\documentclass[journal, twocolum, twoside]{IEEEtran}
\usepackage{amssymb}
\usepackage{amsmath}
\usepackage{amsfonts}
\usepackage[outercaption]{sidecap}
\usepackage{graphicx}%
\usepackage{cite}%
\usepackage{balance}%
\usepackage{anyfontsize}
\newenvironment{rcases}
  {\left\lbrace\begin{aligned}}
  {\end{aligned}\right.}

\title{Coordinated Tethering for Multi-RAT Cellular Networks: An Algorithmic Solution and Performance Analysis }
\vspace{-0.0cm}
\author{{Petros~S.~Bithas,~\IEEEmembership{Member,~IEEE,} and Athanasios S. Lioumpas}

\thanks{P. S. Bithas is with the Department of Digital Systems, University of Piraeus, Greece (e-mail:pbithas@ieee.org).}
\thanks{A. S. Lioumpas is with Cyta Hellas Telecommunications S.A., Athens, Greece (e-mail: athanasios.lioumpas@hq.cyta.gr)}
\thanks{Part of this paper has been presented at IEEE CAMAD}
}
\begin{document}

\maketitle%

\begin{abstract}%
The radio resources management in multi-radio access technology (multi-RAT) heterogeneous networks plays a dominant role in satisfying the high data rates of current and future applications. With the WiFi technology being the spearhead of the wireless local area networks (WLAN)s, the exploitation of already deployed WLANs (e.g., WiFi access points (AP)s) has attracted considerable attention, as an efficient and practical method to improve the performance of wireless networks. Based on the coordinated tethering concept, we introduce a purely wireless heterogeneous network deployment, where cellular and WLAN radio resources are optimally coordinated towards the universal maximization of user throughput. The wireless users are instructed by the eNB about their role in the network (normal user or AP) and the access technology they have to employ. Important performance metrics of the proposed hybrid scheme, including the bit error probability, the ergodic capacity and the average signal-to-interference-plus noise ratio (SINR), are theoretically studied and closed form expressions are derived for the single-user case with multiple interferers, for both identical and non-identical fading conditions. Also, based on the general multi-cellular hybrid WLAN/Cellular concept, we first propose an intercell interference minimization approach. Then we present a novel scheme for achieving frequency reuse equal to one within a single macro-cell, under specific performance criteria and constraints, that guarantee the overall cell's or the individual user's quality-of-service requirements. For the latter case we consider two optimization problems that aim at the overall cell's SINR maximization or at the minimum user's SINR maximization. For all optimization problems, we propose fast greedy solutions. Numerical results and simulations show that the proposed wireless architecture may offer significant performance gains in the presence of multiple interferers, compared to a conventional cellular network.
\end{abstract}%
\markboth{submitted to Transactions on Emerging Telecommunications Technologies}{submitted to Transactions on Emerging Telecommunications Technologies}
\vspace{-0.0cm}

\section{Introduction}
The demand for wireless data traffic is growing faster than the capacity provided by the network, due to the highly increased number of different type of connected devices and the nature of the applications, which are getting more bandwidth demanding (e.g., video streaming). With the link efficiency (e.g., advanced physical layer techniques) having reached its performance limits and the increase of licensed spectrum being impractical, it has been well established that further spectral efficiency improvements are only possible by increasing the node deployment density \cite{Femto3}, or by utilizing the unlicensed spectrum bands in an intelligent way.

The two major approaches towards denser network deployments are: a) network splitting into smaller macro-cells, and b) adoption of the Heterogeneous Networks (HetNet)s and multi-radio access technology (multi-RAT) network concepts. The cell splitting may not always be an efficient solution, especially in already dense deployments, where the additional intercell interference is prohibitive \cite{Femto3}. Under the cooperative framework, the concept of HetNets, relies on the deployment of heterogeneous low power nodes (LPN)s within the macro-cell \cite{Duong1_cognitive,Himal_cooperative,Sami_cognitive,Maged_cognitive}. The HetNet deployments provide a wide area coverage through the macro cell and a more targeted coverage of special zones (e.g., areas with increased traffic or areas with weak signal reception) through the LPNs \cite{Femto1,Hetnet1,Li_Heterogeneous,Shakir_femto,Mukherjee_Heterogeneous, Haijun_1,Haijun_2,Haijun_3,Haijun_4}, which may use a different access technology than the macro cell, e.g., WiFi. The HetNet concept is also part of the 3rd Generation Partnership Project (3GPP) Long Term Evolution (LTE) network architecture, where the LPNs include the picocells, femtocells, home eNodeBs (eNB)s and relay nodes \cite{LTE_femto}. Furthermore, the utilization of the WLANs for data and voice over IP applications is also part of the 3GPP LTE specifications \cite{WLAN}.

\subsection{Motivation and related works}

The capacity offloading to non-cellular radio technologies, especially to 802.11-based wireless local area networks (WLAN)s (i.e., WiFi), has attracted considerable research interest as a cost-efficient, easy to deploy solution \cite{Wifi_off2}, \cite{Wifi_off}. This offloading is particularly attractive for applications that generate delay-tolerant data, e.g., data generated in wireless sensor networks, machine-to-machine (M2M) applications or video downloads. Building more WiFi hot spots is significantly more cost efficient than network upgrades or small-cells deployments \cite{Wifi_off3}. Furthermore, taking into account the huge number of WiFi access points (AP)s already installed at home or at work, it becomes evident that a very dense network is already deployed. It is interesting to note that the IEEE 802.11 standard includes a convergence with 3GPP standards through the Extensible Authentication Protocol-Subscriber Identity Module (EAP-SIM) protocol for authentication and key agreement protocol \cite{Wifi_off4}, \cite{Wifi_off5}, which is an enabler for utilizing the WLAN APs for offloading cellular data in practice.

More recently, the concept of tethering was employed in \cite{Teth1}, \cite{Teth2}, \cite{Teth3} so that a single BS in the cell can
serve more users with the help of hotspots utilizing the television white spaces (TVWS) frequencies. The authors proposed a clustering algorithm and investigated different handover scenarios. The authors are employing the tethering technology in order to provide a solution, which does not require additional infrastructure, in contrast to the small-cells deployments, or network splitting. However, the devices need to be capable of operating on both LTE and TVWS bands with the same cellular access technology.

 The offloading through opportunistic communications and social participation for mobile data generated by mobile social networks (MoSoNets), was introduced in \cite{SocialNet}, \cite{SocialNet2}. According to this scheme the service providers may deliver the information to specific target-users. Then, these users propagate the information among all the subscribed users if their mobile phones are within the transmission range of each other via WiFi or Bluetooth. Despite the limitation of this scheme to broadcast information in MoSoNets, it demonstrates that the connection sharing may reduce the traffic load through the cellular network. In addition to offloading schemes, where the WLAN APs employ a wired backhaul, e.g., asymmetric digital subscriber line (ADSL), in \cite{Tether} the authors presented an offloading scheme based on WiFi Tethering, where the cellular radio links of one or more mobile smartphones are exploited for building WiFi hotspot, which can be then used by vicinal mobile users. Similar network setups have been also studied under the framework of cooperative relay-assisted network. For example \cite{cooperative_kim} investigates the performance of a multicell uplink relay network, where each cell contains a source, a destination, and multiple relays, assuming that multiple interferers affect only the relays. The authors in \cite{cooperative_aissa} study the effect of co-channel interferers on outage probability (OP) of the secondary link in a spectrum-sharing environment, where pairs of secondary users communicate through a relay.
\begin{figure*}[t!]
\centering
\includegraphics[keepaspectratio,width=12.5cm, trim=0.0cm 0.0cm 0cm 2cm]{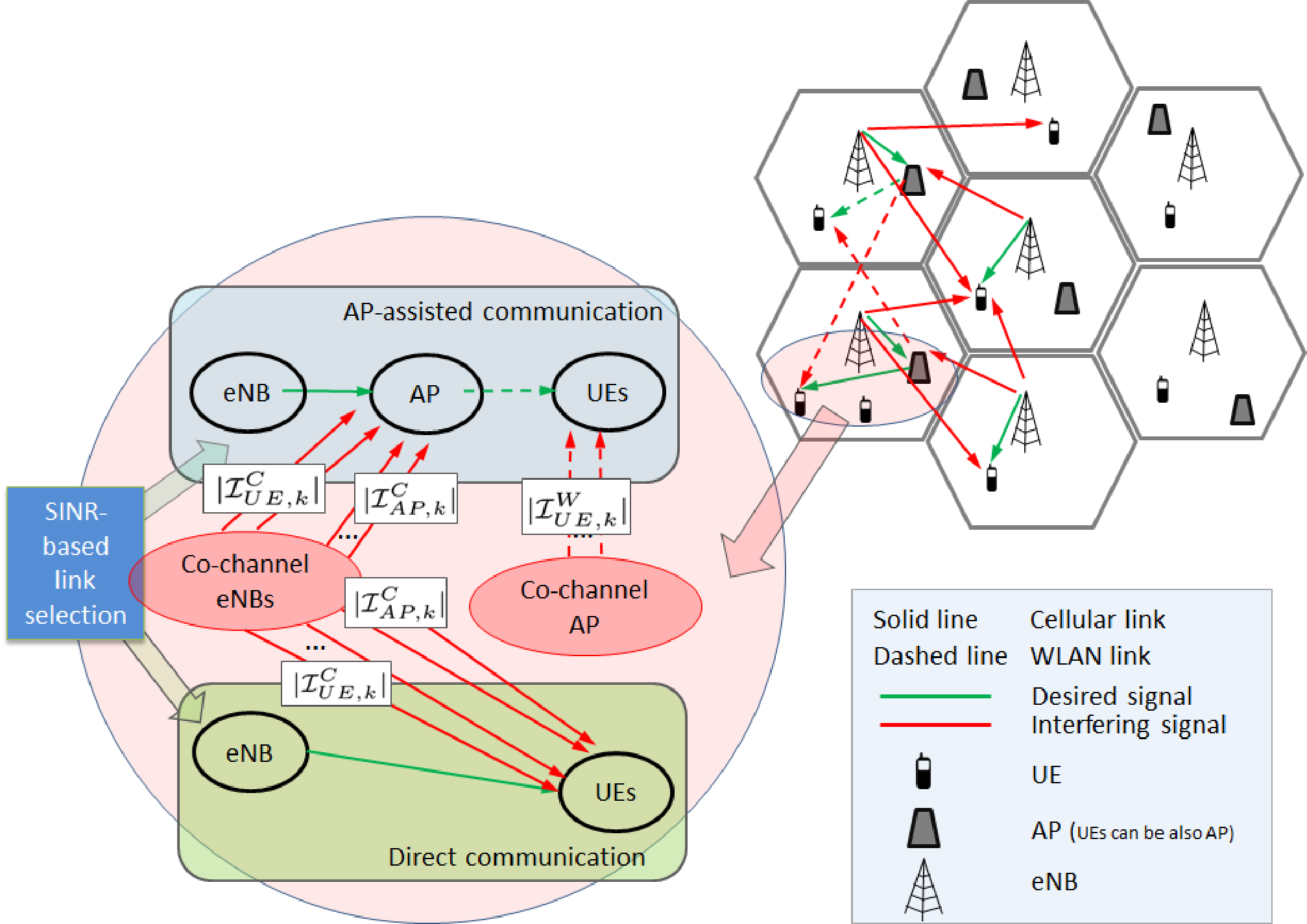}
\caption{The hybrid Cellular/WLAN architecture and interference model.} \label{Fig:Fig2a}
\end{figure*}

\subsection{Contribution}
In this paper, we employ a purely wireless hybrid cellular/WLAN communication approach, where the mobile users can be served by either the eNB or a WLAN AP, depending on the selection strategy (e.g., based on the measured signal to interference plus noise ratio (SINR)). In contrast to the conventional offloading approach, where the WLAN APs have a wired backhaul, a tethering approach is adopted. According to this scheme the WLAN APs are wirelessly connected to the eNB and share this broadband connection with specific users over WLAN frequencies. Then the users select their serving node, i.e., the macro-cell eNB or a WLAN AP, based on a performance criterion. The aim of this architecture is to reduce the transmission power from the eNB to users with bad channel conditions (e.g., users located at the cell edge) and thus the total interference at both the Cellular and WLAN part of the hybrid network, while avoiding any modifications to the existing cellular network. In contrast to previous works (e.g., \cite{Teth1}, \cite{Teth2}, \cite{Teth3}), we assume that conventional WiFi technology is used, which implies that regular devices can be utilized. Furthermore, we consider a holistic approach where the total cell throughput is maximized, meaning that the eNB coordinates both the cellular and WLAN resources allocation, as well the orle of each device.

We provide a complete theoretical performance analysis, taking into account important performance metrics of the proposed hybrid scheme, namely the average bit error probability (ABEP), the average output SINR, the ergodic capacity and the OP. Closed form expressions are derived for the single-user case with multiple interferers, for both identical and non-identical Rayleigh fading channels.

Additionally, the hybrid scheme under consideration is extended to a multi-user (MU) communication scenario. Specifically, considering a multi-cell network scenario and based on the proposed approach, the system performance in optimized, by minimizing intercell interference. This optimization has been obtained based on exhaustive search as well as via a fast greedy algorithm that provides solutions close to the optimal ones. Furthermore, for the general multi-cellular hybrid Cellular/WLAN concept, we present a novel scheme for achieving frequency reuse one within a single macro-cell, under specific performance criteria and constraints that guarantee the overall cell's or the individual user's quality-of-service (QoS) requirements. For the latter case we consider two optimization problems namely, i) maximization of the overall cell's SINR and ii) maximization of the minimum user's SINR for the sake of fairness among users. For these problems, two fast greedy algorithm are proposed, which provide suboptimal solutions close to the optimal ones, obtained via exhaustive search. Numerical results showed that the proposed wireless architecture may offer significant performance gains in the presence of multiple interferers, compared to a conventional cellular network.

\subsection{Paper structure}

The rest of the paper is organized as follows. Section II introduces the system and interference model, Section III presents analytical results for the single user case with multiple interferers. The multi-cell MU performance optimization is discussed in Section IV, while in Section V the single cell frequency reuse equal to one optimization problem is considered. Section VI presents some numerical results that demonstrate the benefits of the proposed architecture and Section VII includes some conclusion remarks. For the reader's convenience, most of the notations and symbols used in the rest of the paper are summarized in Table I.

\begin{table}[b!]
\renewcommand{\arraystretch}{1.1}
\caption {Notation and Symbols}
\label{Tab:1} \centering
\begin{tabular}{|r|l|}
  \hline
  $z_{eu_{\textit{\scriptsize  j,i}}}$ & Complex channel gain between the $j$th eNB and the $i$th UE \\
  \hline
  $z_{ea_{\textit{\scriptsize  j,i}}}$ & Complex channel gain between the $j$th eNB and the $i$th AP \\
  \hline
   $z_{au_{\textit{\scriptsize  j,i}}}$ & Complex channel gain between the $j$th AP and the $i$th UE\\
  \hline
  $s_{eu_{\textit{\scriptsize  j,i}}}$ & The complex symbol transmitted by the $j$th eNB, \\
             & targeting the $i$th UE \\
  \hline
  $s_{ea_{\textit{\scriptsize  j,i}}}$ &  The complex symbol transmitted by the $j$th eNB, \\
             & targeting the $i$th AP \\
  \hline
   $s_{au_{\textit{\scriptsize  j,i}}}$ &  The complex symbol transmitted by the $j$th AP,  \\
                  & targeting the $i$th UE \\
  \hline
  $\mathcal{I}_{UE,i}^{C}$ & The set of interfering UEs using the same cellular frequency  \\
                           & as the $i$th UE or AP \\
  \hline
  $\mathcal{I}_{AP,i}^{C}$ & The set of interfering APs using the same cellular frequency \\
                           & as the $i$th UE or AP  \\
  \hline
  $\mathcal{I}_{UE,i}^{W}$ & The set of interfering UEs using the same WLAN frequency \\
                           & as the $i$th UE  \\
  \hline
   $z_{I_{eu,j}}$ & Complex channel gain of the $jth$ UE $\in \mathcal{I}_{UE,i}^{C}$ \\
  \hline
  $z_{I_{au,j}}$ & Complex channel gain of the $jth$ AP $\in \mathcal{I}_{AP,i}^{C}$ \\
  \hline
  $z_{I_{ea,j}}$ & Complex channel gain of the $jth$ UE $\in \mathcal{I}_{UE,i}^{W}$ \\
  \hline
  $s_{I_{eu,j}}$ &  The complex symbol transmitted by an eNB, targeting the $jth$\\
               & UE $\in \mathcal{I}_{UE,i}^{C}$ \\
  \hline
  $s_{I_{ea,j}}$ &  The complex symbol transmitted by an eNB, targeting the $jth$\\
               & AP $\in \mathcal{I}_{AP,i}^{C}$ \\
  \hline
   $s_{I_{au,j}}$ &  The complex symbol transmitted by an AP, targeting the $jth$\\
                    & UE $\in \mathcal{I}_{UE,i}^{W}$ \\
  \hline
  $P_{eu_{\textit{\scriptsize  j,i}}}$ & The power of the complex symbol transmitted by the $j$th eNB, \\
               & targeting the $i$th UE, $P_{eu_{\textit{\scriptsize  j,i}}}=\mathbb{E}\left< |s_{eu_{\textit{\scriptsize  j,i}}}|^2\right>$ \\
  \hline
  $P_{ea_{\textit{\scriptsize  j,i}}}$ & The power of the complex symbol transmitted by the $j$th eNB, \\
              & targeting the $i$th AP, $P_{ea_{\textit{\scriptsize  j,i}}}=\mathbb{E}\left< |s_{ea_{\textit{\scriptsize  j,i}}}|^2\right>$ \\
  \hline
   $P_{au_{\textit{\scriptsize  j,i}}}$ & The power of the complex symbol transmitted by the $j$th AP,  \\
                   & targeting the $i$th UE, $P_{au_{\textit{\scriptsize  j,i}}}=\mathbb{E}\left< |s_{au_{\textit{\scriptsize  j,i}}}|^2\right>$ \\
  \hline
  $P_{I_{eu,j}}$ & The power of the complex symbol transmitted by an eNB, \\
               & targeting the $jth$ UE $\in \mathcal{I}_{UE,i}^{C}$, $P_{I_{eu,j}}=\mathbb{E}\left< |s_{I_{eu,j}}|^2\right>$  \\
  \hline
  $P_{I_{ea,j}}$ & The power of the complex symbol transmitted by the eNB, \\
               & targeting the $jth$ AP $\in \mathcal{I}_{AP,i}^{C}$, $P_{I_{ea,j}}=\mathbb{E}\left< |s_{I_{ea,j}}|^2\right>$ \\
  \hline
   $P_{I_{au,j}}$ & The power of the complex symbol transmitted by an AP, \\
                    & targeting the $jth$ UE $\in \mathcal{I}_{UE,i}^{W}$, $P_{I_{au,j}}=\mathbb{E}\left< |s_{I_{au,j}}|^2\right>$ \\
  \hline
  $\mathbb{E}\left< \cdot\right>$ &  The statistical averaging\\
  \hline
  $|\mathcal{A}|$ & The size of a set $\mathcal{A}$ \\
  \hline
\end{tabular}
\end{table}

\section{System and Interference Model}
We consider the downlink of a multi-RAT, multi-cell wireless network, where the macro-cells are underlaid with $M$ wireless APs, which are not connected to a wired backhaul. Because of the frequency reuse pattern among the macrocells, the user equipment (UE)s and the APs in each macrocell are subject to co-channel interference due to a number of nodes. The APs as well as the UEs are equipped with two radio access interfaces (RAI)s, namely a cellular RAI (e.g., LTE) and a WLAN (e.g., WiFi) RAI, each operating at a different frequency band. Depending on the communication strategy (e.g., based on the received SINR), each UE might be served either through:

 \begin{enumerate}
  \item a direct communication link to the eNB (one phase communication, eNB$\rightarrow$UE ) utilizing cellular frequencies, or
  \item an indirect link to the eNB via an AP, which decodes and forwards the received signal to the target UE\footnote{The APs may be non-user nodes, dedicated for forwarding the data from the eNB to the UEs, or idle UEs, which can be used occasionally as APs. The type of relaying at the AP does not affect the scope and the conclusions of this work.} (two-phase communication, eNB$\rightarrow$AP$\rightarrow$UE), utilizing cellular frequencies for the eNB$\rightarrow$AP link and WLAN frequencies for the AP$\rightarrow$UEs link\footnote{It is noted that the WLAN is assumed to be operating in the non-saturation regime and impairments due to the contention between UEs at the WLAN AP are not taken into consideration \cite{alexei}.}.
\end{enumerate}

A UE, that is directly connected to the eNB, will experience interference caused by adjacent eNBs, which serve other UEs and APs at the same cellular frequencies. On the other hand, considering the eNB$\rightarrow$AP$\rightarrow$UE scenario, in the first phase, the AP will experience interference from adjacent serving other UEs and APs at the same cellular frequencies. In the second phase the UE experiences the interference from other APs that serve UEs at the same WLAN frequencies.

Under these assumptions, the complex baseband signals transmitted by the $n$th eNB and received by the $k$th UE can be expressed as
\begin{equation}\label{eq:received_signals1}
\begin{split}
y_{{eu_{\textit{\scriptsize n,k}}}} =&z_{{eu_{\textit{\scriptsize n,k}}}} s_{{eu_{\textit{\scriptsize n,k}}}} + \sum \limits_{i=1}^ {|\mathcal{I}_{UE,k}^{C}|}  z_{I_{eu,i}}s_{I_{eu,i}} \hspace{0.4cm}\\ &  +  \sum\limits_{i=1}^{|\mathcal{I}_{AP,k}^{C}|} z_{I_{ea,i}}s_{I_{ea,i}} + w_{{eu,k}}.
\end{split}
\end{equation}
Moreover, the complex baseband signal transmitted by the $m$th AP and received by the $k$th UE is
\begin{IEEEeqnarray}{rCl}\label{eq:received_signals3}
y_{{au_{\textit{\scriptsize m,k}}}}   = z_{{au_{\textit{\scriptsize m,k}}}}s_{{au_{\textit{\scriptsize m,k}}}} +  \sum\limits_{i=1}^{|\mathcal{I}_{UE, k}^{W}|} z_{I_{au,i}}s_{I_{au,i}}  + w_{{au,k}}
\end{IEEEeqnarray}
and that transmitted by the $n$th eNB and received by the $k$th AP
\begin{equation}\label{eq:received_signals2}
\begin{split}
y_{{ea_{\textit{\scriptsize n,k}}}}=& z_{{ea_{\textit{\scriptsize n,k}}}}s_{{ea_{\textit{\scriptsize n,k}}}} +  \sum\limits_{i=1}^{|\mathcal{I}_{UE,k}^{C}|} z_{I_{eu,i}}s_{I_{eu,i}}
\\ & \hspace{0.4cm}+ \sum\limits_{i=1}^{|\mathcal{I}_{AP,k}^{C}|} z_{I_{ea,i}}s_{I_{ea,i}} +w_{{ea,k}}.
\end{split}
\end{equation}
In these equations $w_{{eu,k}}$, $w_{{au,k}}$ and $w_{{ea,k}}$ denote the complex additive white Gaussian noise (AWGN) with zero mean and variance $N_0$ at the corresponding UEs and APs.

It is noted that similar to \cite{6327373}, the channel gains of both desired and interfering signals are related to the distance as well as the propagation path loss between the transmitter and the receiver. Next we assume that the envelopes of channel complex gains, i.e., $h_{{eu_{\textit{\scriptsize n,k}}}}$, $h_{{ea_{\textit{\scriptsize n,k}}}}$, $h_{{au_{\textit{\scriptsize m,k}}}}$, $h_{I_{eu,i}}$, $h_{I_{au,i}}$, $h_{I_{ea,i}}$, follow the Rayleigh distribution and hence their squares are exponentially distributed.

\subsection{Interference in a Conventional Cellular Network}
Considering the case of a conventional cellular network, where the frequency resources are reused among macrocells, the instantaneous SINR of the $k$th UE served by the $n$th eNB can be expressed as
\begin{equation} \label{eq:SINR_cell}
\gamma_{{c,k}}=\frac{\gamma_{eu_{\textit{\scriptsize n,k}}}}{1+\underbrace{\sum\limits_{i=1, i \neq k}^{|\mathcal{I}_{UE,k}^{C} |}\gamma_{I_{eu,i}}}_{\rm eNB \rightarrow UEs}}
\end{equation}
where $\gamma_{eu_{\textit{\scriptsize n,k}}}=h_{eu_{\textit{\scriptsize n,k}}}^2 P_{eu_{\textit{\scriptsize n,k}}}/N_0$ is the instantaneous received signal-to-noise ratio (SNR) of the $k$th UE and $\gamma_{I_{eu,i}}=h_{I_{eu,i}}^2 P_{I_{eu,i}}/N_0$ is the interference-to-noise ratio (INR) due to the cellular co-channel interferers, which follows the exponential distribution. As mentioned above, $\gamma_{eu_{\textit{\scriptsize n,k}}}$ and $\gamma_{I_{eu,i}}$ are exponentially distributed random variables (RV)s with mean values $\overline{\gamma}_{eu_{\textit{\scriptsize n,k}}}=\mathbb{E}\left<h_{eu_{\textit{\scriptsize n,k}}}^2\right> P_{eu_{\textit{\scriptsize n,k}}}/N_0$ and $\overline{\gamma}_{I_{eu,i}}=\mathbb{E}\left<h_{I_{eu,i}}^2\right> P_{I_{eu,i}}/N_0$, respectively.

\subsection{Interference in the Hybrid Cellular/WLAN Network}
Regarding the hybrid Cellular/WLAN network under investigation and the case where the UE is directly connected with the $n$th eNB (employing the cellular RAI), the instantaneous SINR of the $k$th UE can be expressed as
\begin{equation}\label{eq:SINR_hybrid_direct}
\gamma_{{{eu,k}}}=\frac{\gamma_{eu_{\textit{\scriptsize n,k}}}}{1+\underbrace{\sum\limits_{i=1}^{|\mathcal{I}_{AP,k}^{C}|} \gamma_{I_{ea,i}}}_{\rm eNB \rightarrow AP}+\underbrace{\sum\limits_{i=1}^{|\mathcal{I}_{UE,k}^{C} |}\gamma_{I_{eu,i}}}_{\rm eNB \rightarrow UE}}
\end{equation}
where $\gamma_{I_{ea,i}}=h_{I_{ea,i}}^2 P_{I_{ea,i}}/N_0$ is the INR due to the cellular co-channel interferers, which follows the exponential distribution with mean values $\overline{\gamma}_{I_{ea,i}}=\mathbb{E}\left<h_{I_{ea,i}}^2\right> P_{I_{ea,i}}/N_0$.

Considering the case where the target UE is connected to a eNB via a WLAN AP, in the first phase of the transmission the AP will experience interference from the eNB serving other UEs and the APs at the same cellular frequencies. In the second phase the UE experiences the interference from the APs that serve UEs at the same WLAN frequencies. In this sense, during the first phase, the instantaneous SINR at the $m$th AP, directly connected with the $n$th eNB, will be
\begin{equation}\label{eq:SINR_hybrid_indirect_EA}
\gamma_{ ea,m}=\frac{\gamma_{ea_{\textit{\scriptsize n,m}}}}{1+\underbrace{\sum\limits_{i=1}^{|\mathcal{I}_{AP,m}^{C}|} \gamma_{I_{ea,i}}}_{\textrm eNB \rightarrow AP}+\underbrace{\sum\limits_{i=1}^{| \mathcal{I}_{UE,m}^{C}|}\gamma_{I_{eu,i}}}_{\textrm eNB \rightarrow UE}}
\end{equation}
where $\gamma_{ea_{\textit{\scriptsize n,m}}}=h_{ea_{\textit{\scriptsize n,m}}}^2 P_{ea_{\textit{\scriptsize n,m}}}/N_0$ is the instantaneous SNR at the $m$th AP, following the exponential distribution with mean value $\overline{\gamma}_{{ea_{\textit{\scriptsize n,m}}}}=\mathbb{E}\left<h_{ea_{\textit{\scriptsize n,m}}}^2\right> P_{ea_{\textit{\scriptsize n,m}}}/N_0$. In the second phase of the communication, where the UE receives the desired signal from one AP and interfering signals coming from the $m$th APs, the instantaneous SINR at the $k$th UE can be expressed as
\begin{equation}\label{eq:SINR_hybrid_indirect_AE}
\gamma_{ a u,k}=\frac{\gamma_{au_{\textit{\scriptsize m,k}}}}{1+\underbrace{\sum\limits_{i=1}^{|\mathcal{I}_{UE,k}^{W}|} \gamma_{I_{au,i}}}_{\textrm AP \rightarrow UE}}
\end{equation}
where $|\mathcal{I}_{UE,m}^{C}|+|\mathcal{I}_{UE,k}^{W}|=N$, $\gamma_{au_{\textit{\scriptsize m,k}}}=h_{au_{\textit{\scriptsize m,k}}}^2 P_{au_{\textit{\scriptsize m,k}}}/N_0$ is the instantaneous SNR at the $k$th UE and $\gamma_{I_{au,i}}=h_{I_{au,i}}^2 P_{I_{au,i}}/N_0$ is the INR due to WLAN co-channel interferers. Both $\gamma_{{au_{\textit{\scriptsize m,k}}}}$ and $\gamma_{I_{au,i}}$ are exponentially distributed with mean values $\overline{\gamma}_{au_{\textit{\scriptsize m,k}}}=\mathbb{E}\left<h_{{au_{\textit{\scriptsize m,k}}}}^2\right> P_{au_{\textit{\scriptsize m,k}}}/N_0$ and $\overline{\gamma}_{I_{au,i}}=\mathbb{E}\left<h_{I_{au,i}}^2\right> P_{I_{au,i}}/N_0$, respectively. Note that the sets $\mathcal{I}_{UE,i}^{C}$ and $\mathcal{I}_{AP,i}^{C}$ do not include the same UEs and APs when considering the direct or indirect communication scenarios.

%For end-to-end performance, the allocation of power is an important factor because the received SINR is proportional to
%the transmitted power. Without losing the generality of our approach, and following a similar assumption as in \cite{6340378,Liang}, we assume that the source and all the relays transmit at the same power (equal power allocation).

\subsection{Link Selection Strategy}

The communication mode of each UE (i.e., direct communication with the eNB or via an WLAN AP) is determined by the eNB, which has a full knowledge of the channel state information for all the links within the cell. In this work, we assume that this decision is based on the SINR, namely the UE connects to that node (eNB or WLAN AP) that results in the maximum individual instantaneous SINR. The signalling details for realizing this scenario is beyond the scope of the paper.

\section{Single User Analysis}
In this Section, important statistical metrics of the instantaneous SINR, namely the probability density function (PDF) and the cumulative distribution function (CDF), are presented for the conventional cellular and the hybrid Cellular/WLAN networks. The derivation procedures for these metrics can be found in Appendices A and B.
\subsection{Statistical Analysis}
\subsubsection{SINR Statistics for the Conventional Cellular Network}\label{sub:eNB_UEs}
Let us consider the conventional cellular network and the SINR at the $k$th UE as given by \eqref{eq:SINR_cell}. In this study it is assumed that the co-channel interfering signals add up incoherently
since this represents a more realistic assessment of the co-channel interference in cellular systems \cite{00966071}. The total instantaneous INR caused by other eNBs to the $k$th UE, is given by
\begin{equation}\label{eq:total_interference_UE_definition}
\gamma_{c_{I,k}}=\sum\limits_{i=1}^{|\mathcal{I}_{UE,k}^C|} \gamma_{I_{eu,i}}.
\end{equation}

Assuming independent but non identical distributed (i.n.d.) fading conditions, the PDF and the CDF of $\gamma_{c,k}$ can be expressed as in the form-A given in Table~II, by substituting $C=\gamma_{c,k}$, $X=|\mathcal{I}_{UE,k}^C|$, $Y=\overline{\gamma}_{{eu_{\textit{\scriptsize n,k}}}}$ and $Z_i=\overline{\gamma}_{I_{eu,i}}$. If identical distributed (i.i.d.) fading is assumed, the PDF and the CDF of $\gamma_{c,k}$ is of the form-A given in Table~III, by substituting $C=\gamma_{c,k}$,  $X=|\mathcal{I}_{UE,k}^C|$, $Y=\overline{\gamma}_{{eu_{\textit{\scriptsize n,k}}}}$ and $Z=\overline{\gamma}_{I_{eu,k}}$.
\begin{table*}
\renewcommand{\arraystretch}{1.6}
\caption{Non Identical Statistics.}
\label{Tab:ABEP_accuracy} \centering
\begin{tabular}{c || c || c}
  \hline\hline
    \begin{tabular}{c}
                    form A
\\
\\
\\
       \hline   \hline
    \\
    \\
    form B

    \\
    \\
    \\
    \\
    \\
    \\

        \end{tabular}
             & \begin{tabular}{c}
                                 PDF \\
             \hline
           \\ CDF  \\ \\
            \hline \hline
                \\    \\  PDF   \\ \\ \\ \\
             \hline
          \\  CDF \\ \\
            \end{tabular}
                                                & \begin{tabular}{c}
                        $ f_{C}(\gamma)= \left( \prod\limits_{i=1}^{X}\frac{1}{Z_i}\right) \sum\limits_{j=1}^{X}  \frac{\exp \left( -\frac{\gamma}{Y}\right)}{\prod\limits_{\substack{l=1 \\ l\neq j}}^{X} \left( \frac{1}{Z_l}-\frac{1}{Z_j}\right)} \frac1{Y}\left[\frac{1}{\left( \frac{1}{Z_j}+\frac{\gamma}{Y}\right)}+\frac{1}{\left( \frac{1}{Z_j}+\frac{\gamma}{Y}\right)^2}\right] $  \\
                                    \hline
                       $F_{C}(\gamma)= 1-\left( \prod\limits_{i=1}^{X}\frac{1}{Z_i}\right) \sum\limits_{j=1}^{X}  \frac{\exp \left( -\frac{\gamma}{\overline{\gamma}_{_{eu,l}}}\right)}{\prod\limits_{\substack{l=1 \\ l\neq j}}^{X} \left( \frac{1}{Z_l}-\frac{1}{Z_j}\right)}  \frac{1}{\left( \frac{1}{Z_j}+\frac{\gamma}{Y}\right)}$  \\
                                               \hline \hline
$f_{C}(\gamma)= \frac{1}{Y}\Bigg\{\sum\limits_{j_1=1}^{X_1}  \sum\limits_{j_2=1}^{X_2} \frac{\left( \prod\limits_{i=1}^{X_1}\frac{1}{Z_{1,i}}\right)\left( \prod\limits_{i=1}^{X_2}\frac{1}{Z_{2,i}}\right)\frac{Z_{1,j_1} Z_{2,j_2}}{Z_{1,j_1}-Z_{2,j_2}}}{\prod\limits_{\substack{k_1=1 \\ k_1\neq j_1}}^{X_1} \left( \frac{1}{Z_{1,k_1}}-\frac{1}{Z_{1,j_1}}\right)\prod\limits_{\substack{k_2=1 \\ k_2\neq j_2}}^{X_2} \left( \frac{1}{Z_{2,k_2}}-\frac{1}{Z_{2,j_2}}\right)}
\left[\frac{\exp \left( -\frac{\gamma}{Y}\right)}{\frac{\gamma}{Y} +\frac{1}{Z_{2,j_1}}} -\frac{\exp \left( -\frac{\gamma}{Y}\right)}{\frac{\gamma}{Y} +\frac{1}{Z_{2,j_2}}}\right]$\\+$\sum\limits_{j_1=1}^{X_1}\sum\limits_{j_2=1}^{X_2} \frac{\left( \prod\limits_{i=1}^{X_1}\frac{1}{Z_{1,i}}\right)\left( \prod\limits_{i=1}^{X_2}\frac{1}{Z_{2,i}}\right)\frac{Z_{2,j_1} Z_{2,j_2}}{Z_{2,j_1}-Z_{2,j_2}}}{\prod\limits_{\substack{k_1=1 \\ k_1\neq j_1}}^{X_1} \left( \frac{1}{Z_{2,k_1}}-\frac{1}{Z_{1,j_1}}\right)\prod\limits_{\substack{k_2=1 \\ k_2\neq j_2}}^{X_2} \left( \frac{1}{Z_{2,k_2}}-\frac{1}{Z_{2,j_2}}\right)}
\left[\frac{\exp \left( -\frac{\gamma}{Y}\right)}{\left(\frac{\gamma}{Y} +\frac{1}{Z_{1,j_1}}\right)^2} -\frac{\exp \left( -\frac{\gamma}{Y}\right)}{\left(\frac{\gamma}{Y} +\frac{1}{Z_{2,j_2}}\right)^2}\right]\Bigg\}$  \\
                                                \hline
                                                $F_{C}(\gamma)= 1- \sum\limits_{j_1=1}^{X_1}\sum\limits_{j_2=1}^{X_2} \frac{\left( \prod_{i=1}^{X_1}\frac{1}{Z_{1,i}}\right)}{\prod_{\substack{k_1=1 \\ k_1\neq j_1}}^{X_1} \left( \frac{1}{Z_{1,k_1}}-\frac{1}{Z_{1,j_1}}\right)}
   \frac{\left( \prod_{i=1}^{X_2}\frac{1}{Z_{2,i}}\right)}{\prod_{\substack{k_2=1 \\ k_2\neq j_2}}^{X_2} \left( \frac{1}{Z_{2,k_2}}-\frac{1}{Z_{2,j_2}}\right)}$ \\ $\times \frac{Z_{1,j_1} Z_{2,j_2}}{Z_{1,j_1}-Z_{2,j_2}}
\left[\frac{\exp \left( -\frac{\gamma}{Y}\right)}{\frac{\gamma}{Y} +\frac{1}{Z_{1,j_1}}} -\frac{\exp \left( -\frac{\gamma}{Y}\right)}{\frac{\gamma}{Y} +\frac{1}{Z_{2,j_2}}}\right]$  \\
                                                \end{tabular}\\

\hline \hline
\end{tabular}
\end{table*}

\subsubsection{SINR Statistics for the Hybrid Cellular/WLAN Network}
For the hybrid Cellular/WLAN network, the following two complementary communication cases are investigated, namely one-phase direct eNB$\rightarrow$UE communication and two-phase indirect eNB$\rightarrow$AP$\rightarrow$UE communication. In the next subsections, analytical expressions for the PDF and the CDF of the AP and UE output SINR will be presented.
\paragraph{Direct Communication}\label{sub:hybrid_direct}
As far as the case where the target UE is directly connected with the eNB is considered, the SINR at the target UE is given by \eqref{eq:SINR_hybrid_direct}. The PDFs of the total instantaneous INR caused to the $k$th UE by the eNB serving UEs at the same frequency, i.e.,
\begin{equation*}
\gamma_{{I_{{eu,k}}}}=\sum_{i=1}^{|\mathcal{I}_{UE,k}^{C}|} \gamma_{I_{eu,i}}
\end{equation*}
and that of the total instantaneous INR caused to the $k$th UE by the eNB serving APs at the same frequency, i.e.,
\begin{equation*}
\gamma_{{I_{ea,k}}}=\sum_{i=1}^{|\mathcal{I}_{AP,k}^{C}|} \gamma_{I_{ea,i}}
\end{equation*}
follow the chi-square distribution. Assuming i.n.d. fading conditions, the PDF and the CDF of $\gamma_{eu,k}$ has the form-B given in Table~II, by substituting $C=\gamma_{eu,k}$, $X_1=|\mathcal{I}_{UE,k}^{C}|$, $X_2=|\mathcal{I}_{AP,k}^{C}|$, $Y=\overline{\gamma}_{{eu_{\textit{\scriptsize n,k}}}}$, $Z_{1,i}=\overline{\gamma}_{I_{eu,i}}$, $Z_{2,i}=\overline{\gamma}_{I_{ea,i}}$. For i.i.d. fading conditions, the PDF and the CDF of $\gamma_{c,k}$ is of the form-B given in Table~III, by substituting $C=\gamma_{eu,k}$, $X_1=|\mathcal{I}_{UE,k}^{C}|$, $X_2=|\mathcal{I}_{AP,k}^{C}|$, $Y=\overline{\gamma}_{{eu_{\textit{\scriptsize n,k}}}}$, $Z_{1}=\overline{\gamma}_{I_{eu,k}}$, $Z_{2}=\overline{\gamma}_{I_{ea,k}}$.

\paragraph{Indirect Communication}
In the case of the indirect connection of the UE to the eNB, there are two communication phases. During the first one (i.e., eNB$ \rightarrow $AP), the SINR at the target AP is given by \eqref{eq:SINR_hybrid_indirect_EA}, while in the second phase one (i.e, AP$ \rightarrow$UE), the instantaneous SINR at the target UE is expressed as in \eqref{eq:SINR_hybrid_indirect_AE}. Assuming i.n.d. fading conditions, for the first phase, the PDF and the CDF of $\gamma_{ea,m}$ is of the form-B given in Table~II, by substituting $C=\gamma_{ea,m}$, $X_1=|\mathcal{I}_{UE,m}^{C}|$, $X_2=|\mathcal{I}_{AP,m}^{C}|$, $Y=\overline{\gamma}_{{ea_{\textit{\scriptsize n,m}}}}$, $Z_{1,i}=\overline{\gamma}_{I_{eu,i}}$, $Z_{2,i}=\overline{\gamma}_{I_{ea,i}}$.
For the second phase, the PDF and the CDF of $\gamma_{au,k}$ is of the form-A given in Table~II, by substituting $C=\gamma_{au,k}$, $X=|\mathcal{I}_{UE,k}^W|$, $Y=\overline{\gamma}_{{au_{\textit{\scriptsize m,k}}}}$ and $Z_i=\overline{\gamma}_{I_{au,i}}$. Assuming i.i.d. fading conditions, for the first phase, the PDF and the CDF of $\gamma_{ea,m}$ is of the form-B given in Table~III, by substituting $C=\gamma_{ea,m}$, $X_1=|\mathcal{I}_{UE,m}^{C}|$, $X_2=|\mathcal{I}_{AP,m}^{C}|$, $Y=\overline{\gamma}_{{ea_{\textit{\scriptsize n,m}}}}$, $Z_{1}=\overline{\gamma}_{I_{eu,m}}$, $Z_{2}=\overline{\gamma}_{I_{ea,m}}$.
For the second phase, the PDF and the CDF of $\gamma_{au,k}$ is of the form-A given in Table~III, by substituting $C=\gamma_{au,k}$, $X=|\mathcal{I}_{UE,k}^W|$, $Y=\overline{\gamma}_{{au_{\textit{\scriptsize m,k}}}}$ and $Z=\overline{\gamma}_{I_{au,k}}$.

\underline{Indirect Communications Total Instantaneous SINR:}
In the case where the AP decodes and forwards the data to the target UE, the instantaneous SINR at the output of the eNB$\rightarrow$APs$\rightarrow$UEs link can be expressed as \cite{Duong1_cognitive}
\begin{equation}\label{eq:instantaneous_SINR_DF}
\gamma_{ eau,k}(\gamma)= \min \left( \gamma_{ea,k},\gamma_{au,k}\right)
\end{equation}
and thus the corresponding CDF of $\gamma_{eau,k}$ can be expressed as
\begin{align}\label{eq:CDF_SINR_DF} \nonumber
&F_{\gamma_{eau,k}}(\gamma)= \textrm {Pr} \left\{ \min\left(\gamma_{ea,k},\gamma_{au,k}\right)<\gamma\right\} \\ \nonumber
 &= 1-\left[ 1-F_{\gamma_{ea,k}}(\gamma)\right] \left[ 1-F_{\gamma_{au,k}}(\gamma)\right]\\
 &=F_{\gamma_{ea,k}}(\gamma)+F_{\gamma_{au,k}}(\gamma)-F_{\gamma_{ea,k}}(\gamma) F_{\gamma_{ea,k}}(\gamma).
\end{align}
Furthermore, the corresponding PDF expression of $\gamma_{eau,k}$ is the following
\begin{equation}\label{eq:PDF_SINR_DF}
\begin{split}
&f_{\gamma_{eau,k}}(\gamma)= f_{\gamma_{ea,k}}(\gamma)+f_{\gamma_{au,k}}(\gamma)  \\ & -f_{\gamma_{ea,k}}(\gamma)F_{\gamma_{au,k}}(\gamma) -f_{\gamma_{au,k}}(\gamma)F_{\gamma_{ea,k}}(\gamma).
\end{split}
\end{equation}
\begin{table*}
\renewcommand{\arraystretch}{1.6}
\caption{Identical Statistics.}
\label{Tab:ABEP_accuracy} \centering
\begin{tabular}{c || c || c}
  \hline\hline
    \begin{tabular}{c}
                    form A
\\ \\
       \hline   \hline
    \\
    \\
    form B

    \\
    \\
    \\
    \\
    \\  \\ \\

        \end{tabular}
             & \begin{tabular}{c}
                                 PDF \\
             \hline
            CDF  \\
            \hline \hline
                      PDF \\ \\ \\ \\
             \hline
            CDF \\ \\ \\ \\ \\
            \end{tabular}
                                                & \begin{tabular}{c}
                        $ f_{C}(\gamma)= \frac{Z^{-X}}{Y}  \frac{\exp \left( -\frac{\gamma}{Y}\right)}{\left( \frac{\gamma}{Y}+\frac1{Z} \right)^{X}} \left( 1+\frac{X}{\frac{\gamma}{Y} +\frac{1}{Z}}\right)$  \\
                                    \hline
                       $F_{C}(\gamma)= 1-\frac1{Z^{X}} \left( \frac{\gamma}{Y}+\frac{1} {Z}\right)^{-X}\exp \left( -\frac{\gamma}{Y}\right)$  \\
                                               \hline \hline
$
f_{C}(\gamma)= \frac{\frac{1}{Y}}{Z_{2,k}^{X_2} \Gamma\left(X_2\right)} \frac{\exp\left( -\gamma/Y\right)}{Z_{{1,k}}^{X_1} \Gamma\left(X_1\right)} \sum\limits_{j=0}^{X_2-1}  \frac{\binom{X_2-1}{j} (-1)^j\Gamma\left( X_1+j\right) }{\left( \frac{1}{Z_{1,k}}-\frac{1}{Z_{1,k}}\right)^{j+X_1}} \left\{
\frac{\Gamma \left( X_2-j\right)}{\left(\frac{\gamma}{Y}+\frac{1}{Z_{1,k}}\right)^{X_2-j}}
 \right.$
 \\ $\left. \times \left( 1+\frac{X_2-j}{ \frac{\gamma}{Y}+\frac{1}{Z_{1,k}}}\right)- \sum\limits_{k=0}^{X_1+j-1} \frac{\Gamma(X_2+k-j)}{k!} \frac{\left( \frac{1}{Z_{{1,k}}}-\frac{1}{Z_{1,k}}\right)^{k}}{\left( \frac{\gamma}{Y}+\frac{1}{Z_{{1,k}}}\right)^{X_2+k-j}} \left( 1+\frac{X_2+k-j}{\frac{\gamma}{Y}+\frac{1}{Z_{{1,k}}}} \right)\right\}.$  \\
                                                \hline
            $F_{C}(\gamma) =\frac1{Z_{1,k}^{X_2} \Gamma\left( X_2\right)} \frac1{Z_{{1,k}}^{X_1} \Gamma\left( X_1\right)} \sum\limits_{j=0}^{X_2-1} \frac{(-1)^j\binom{X_2-1}{j} \Gamma\left(X_1+j\right)}{\left(1/Z_{{1,k}}-1/ Z_{2,k}\right)^{X_1+j}}\left\{ \left[ \frac{\Gamma\left(X_2-j\right)}{Z_{2,k}^{j-X_2}} \right. \right.$ \\ $\left. \left. - \sum\limits_{n=0}^{X_1+j-1} \frac{\left(1/Z_{{1,k}}-1/Z_{2,k}\right)^n}{n!}  \frac{\Gamma\left(X_2+n-j\right)}{Z_{{1,k}}^{j-X_2-n}} \right]  -\exp \left(-\frac{\gamma}{Y} \right) \left[ \frac{\Gamma\left(X_2-j\right)}{\left(\gamma/Y+1/Z_{2,k}\right)^{X_2-j}} \right. \right. $\\ $\left. \left. -\sum\limits_{n=0}^{X_1+j-1} \frac{\left(1/Z_{{1,k}}-1/Z_{1,k}\right)^{n}} {\left(\gamma/Z_{{1,k}}+1/Z_{{1,k}}\right)^{X_2+n-j}} \frac{\Gamma\left(X_2-j+n\right)}{n!}\right]\right\}.$  \\
                        \end{tabular}\\

\hline \hline
\end{tabular}
\end{table*}

\paragraph{Total Hybrid Cellular/WLAN Network Output SINR}\label{par:total_hybrid_statistics}
Since in the mode of operation, (eNB$\rightarrow$UE or eNB$\rightarrow$AP$\rightarrow$UE), the UE selects the communication link to be connected that provides the maximum SINR, the total instantaneous end-to-end received SINR can be finally expressed as
\begin{equation}\label{eq:instantaneous_SINR_E2E}
\gamma_{\textrm{tot,k}}(\gamma)= \max \left( \gamma_{eu,k},\gamma_{eau,k}\right).
\end{equation}
Therefore, the CDF of $\gamma_{\textrm{tot,k}}$ is given by
\begin{equation}\label{eq:CDF_E2E}
F_{\gamma_{\textrm{tot,k}}}(\gamma)=F_{\gamma_{eu,k}}(\gamma)F_{\gamma_{eau,k}}(\gamma)
\end{equation}
while the corresponding PDF expression can be obtained as
\begin{equation}\label{eq:PDF_E2E}
\begin{split}
f_{\gamma_{\textrm{tot,k}}}(\gamma)&=f_{\gamma_{eu,k}}(\gamma)F_{\gamma_{eau,k}}(\gamma) +F_{\gamma_{eu,k}}(\gamma)f_{\gamma_{eau,k}}(\gamma).
\end{split}
\end{equation}

\subsection{Performance Evaluation}
In this Section, assuming i.i.d. fading conditions and using the previous derived expressions for the PDF and the CDF of the output SINR, important performance criteria will be investigated. More specifically, the proposed system performance will be investigated employing ABEP, average output SINR, ergodic capacity and outage probability criteria. It is noted that the analytical framework presented in this Section can also be applied to the i.n.d. fading scenario. However, due to space limitations these results are not presented here.

\subsubsection{Average Bit Error Probability (ABEP)}
For several modulation schemes, the system's ABEP can be evaluated employing the PDF-based approach \cite{karas_squared_nakagami}. In this case,
the ABEP, denoted as $P_{be}$, can be directly expressed as
\begin{equation}\label{eq_BER_out_definition}
P_{be}=A  \int_0^\infty \exp(-B\gamma)f_{\gamma_{\rm tot,k}}(\gamma)d\gamma
\end{equation}
where $(A,B)$ are constants that depend on the type of modulation. Substituting in this definition the PDF expression provided in \ref{sub:eNB_UEs} (for the conventional cellular case) or \ref{par:total_hybrid_statistics} (for the hybrid Cellular/WLAN case), integrals of the following form appear
\begin{equation}\label{eq:BER_integrals_definition}
\begin{split}
\mathcal{T}_{1,i}(\underbrace{a_i,b_i,c_i}_{i\in\{1,3\}})=\int_0^\infty& \exp \left[ -\left(B+\sum_{j=1}^ic_j\right)x\right]\\ & \times \left[\prod_{j=1}^i \left(\frac1{ x +a_j }\right)^{b_j}\right]dx
\end{split}
\end{equation}
where $a_i, c_i \in \mathbb{R}, b_i \in \mathbb{N}$.
By employing partial fraction and using \cite[eq. (3.382/4)]{Ryzhik}, the following closed-form expression for $\mathcal{T}_{1,i}(a_i,b_i,c_i)$ can be derived as
\begin{equation}\label{eq:BER_final}
\begin{split}
&\mathcal{T}_{1,i}(a_i,b_i,c_i)=\sum_{j=1}^i\sum_{h=1}^{b_j} \Xi_{j,h} \left(B+\sum_{j=1}^ic_j\right)^{h-1}  \\ & \times \exp \left[ a_j\left(B+\sum_{j=1}^ic_j\right)\right]\Gamma \left[ 1-h,\left(B+\sum_{j=1}^ic_j\right)\right]
\end{split}
\end{equation}
where
\begin{displaymath}
\begin{split}
\Xi_{j,h}&=\frac{1}{(b_j-h)!} \frac{d^{b_j-h}}{d\gamma^{b_j-h}} \prod_{{\underset{p\neq j}{p=1}}}^{i} \left( x+a_p\right)^{-b_p} \Big|_{x=-a_j}
\end{split}
\end{displaymath}
and $\Gamma(\cdot,\cdot)$ represents the upper incomplete Gamma function \cite[eq. (8.350/1)]{Ryzhik}. Based on the solution provided in \eqref{eq:BER_final}, a simplified expression for the ABEP of the conventional cellular scenario can be expressed as
\begin{equation}\label{eq:BER_cell}
\begin{split}
&P_{be}^{c}=\frac{A}{\overline{\gamma}_{{eu_{\textit{\scriptsize n,k}}}}} \left( \frac{\overline{\gamma}_{{eu_{\textit{\scriptsize n,k}}}}}{\overline{\gamma}_{I_{{eu,k}}}}\right)^{|\mathcal{I}_{UE,k}^C|} \left( \frac1{\overline{\gamma}_{{eu_{\textit{\scriptsize n,k}}}}}+B\right)^{|\mathcal{I}_{UE,k}^C|-1} \\ & \times \exp \left[ \frac{\overline{\gamma}_{{eu_{\textit{\scriptsize n,k}}}}}{\overline{\gamma}_{I_{{eu,k}}}} \left( \frac{1}{\overline{\gamma}_{{eu,k}}} +B\right)\right]  \\ & \times \left\{ \Gamma\left[ 1-|\mathcal{I}_{UE,k}^C|,\frac{\overline{\gamma}_{{eu_{\textit{\scriptsize n,k}}}}}{\overline{\gamma}_{I_{{eu,k}}}}\left( \frac1{\overline{\gamma}_{{eu_{\textit{\scriptsize n,k}}}}}+B\right)\right]  \right. \\ & \left. +\frac{\left( \frac1{\overline{\gamma}_{{eu_{\textit{\scriptsize n,k}}}}}+B \right)}{\left(|\mathcal{I}_{UE,k}^C|\overline{\gamma}_{{eu_{\textit{\scriptsize n,k}}}}\right)^{-1}} \Gamma\left[ -|\mathcal{I}_{UE,k}^C|,\frac{\overline{\gamma}_{{eu_{\textit{\scriptsize n,k}}}}}{\overline{\gamma}_{I_{{eu,k}}}}\left( \frac1{\overline{\gamma}_{{eu_{\textit{\scriptsize n,k}}}}}+B\right)\right]\right\}.
\end{split}
\end{equation}
It should be noted that the previously presented approach can be also directly applied to the hybrid Cellular/WLAN scenario for obtaining the ABEP. However, the extracted expression for the ABEP is not included here due to space limitations.

\subsubsection{Average Output SINR}\label{subs:ASINR_analysis}
The average output SINR is an important performance indicator that is tightly related to the performance metrics of a system, such as the bit error rate and the asymptotic spectral efficiency. It can be obtained as
\begin{equation}\label{eq_SINR_out_definition}
\overline{\gamma}_{\rm tot,k}= \int_0^\infty \gamma f_{\gamma_{\rm tot,k}}(\gamma)d\gamma.
\end{equation}
Substituting in this definition the PDF expression provided in \ref{sub:eNB_UEs} (for the conventional cellular case) or \ref{par:total_hybrid_statistics} (for the hybrid Cellular/WLAN case), integrals of the following form appear
\begin{equation}\label{eq:SINR_integrals_definition}
\begin{split}
\mathcal{T}_{2,i}\left(\underbrace{a_i,b_i,c_i}_{i\in\{1,3\}}\right)=\int_0^\infty& x \exp \left( -\sum_{j=1}^i c_j x\right) \\ & \times \left[\prod_{j=1}^i \left(\frac1{ x +a_j }\right)^{b_j}\right]dx.
\end{split}
\end{equation}
Similar to the ABEP derivation, in order to simplify \eqref{eq:SINR_integrals_definition}, the partial fraction is employed. Following such an approach and using \cite[eq. (2.3.6/9)]{prudnikov_book_vol1}, $\mathcal{T}_{2,i}(a_i,b_i,c_i)$ can be solved in closed form as
\begin{equation}\label{eq:SINR_solution}
\begin{split}
\mathcal{T}_{2,i}(a_i,b_i,c_i)=&\sum_{j=1}^i\sum_{h=1}^{b_j} \Xi_{j,h} a_j^{2-h}  \\ & \times \Psi \left[ 2,3-h, \left(\sum_{j=1}^ic_j\right)\right]
\end{split}
\end{equation}
where $\Psi(\cdot)$ denotes the confluent hypergeometric function \cite[eq. (9.210/2)]{Ryzhik}. Based on the solution provided in \eqref{eq:SINR_solution}, the average output SINR of the conventional cellular scenario can be expressed as
\begin{equation}\label{eq:SINR_final}
\begin{split}
\overline{\gamma}_{\textrm{c,k}}&=\frac{\overline{\gamma}_{{eu_{\textit{\scriptsize n,k}}}}}{\overline{\gamma}_{I_{{eu,k}}}^2} \Psi\left( 2,3-|\mathcal{I}_{UE,k}^C|,\frac1{\overline{\gamma}_{I_{{eu,k}}}}\right)\\ &+|\mathcal{I}_{UE,k}^C| \frac{\overline{\gamma}_{{eu_{\textit{\scriptsize n,k}}}}}{\overline{\gamma}_{I_{{eu,k}}}} \Psi\left( 2,2-|\mathcal{I}_{UE,k}^C|,\frac1{\overline{\gamma}_{I_{{eu,k}}}}\right).
\end{split}
\end{equation}

It should be noted that the previously presented approach can be also directly applied to the hybrid Cellular/WLAN scenario for obtaining the average output SINR. However, the extracted expression for the average output SINR is not included here due to space limitations.

\subsubsection{Ergodic Capacity}\label{sub_capacity_evaluation}
Ergodic capacity is an essential metric to measure the maximum achievable transmission bit rate under which error is recoverable with Shannon's perspective. In our
system, the overall achievable capacity is given by
\begin{equation}\label{eq:ergodic_capacity_definition}
C_{\gamma_{\rm tot,k}}=\frac1{NH}\int_0^\infty \log_2(1+\gamma) f_{\gamma_{\rm tot,k}}(\gamma)d\gamma
\end{equation}
where $NH$ denotes the number of hops. Since exact closed-form expressions for the capacity is impossible to be derived, next two approaches for evaluating it will be presented. The first one is based on a fast converging infinite series and the second one on an upper bound.
\paragraph{Exact Expression}Substituting in this definition the PDF expression provided in \ref{sub:eNB_UEs} (for the conventional cellular case) or \ref{par:total_hybrid_statistics} (for the hybrid Cellular/WLAN case), and using the infinite series representation of the $\log_2(\cdot)$ function, i.e., \cite[eq. (4.1.29)]{Abramowitz}, integrals of the following form appear
\begin{equation}\label{eq:capacity_integrals_definition}
\begin{split}
\mathcal{T}_{3,i}\left(\underbrace{a_i,b_i,c_o}_{i \in \{1,4\}}\right)=\int_0^\infty& x^D \exp \left( -\sum_{j=1}^i c_j x\right) \\ & \times \left[\prod_{j=1}^{i+1} \left(\frac1{ x +a_j }\right)^{b_j}\right]dx.
\end{split}
\end{equation}
Employing the partial fraction and using \cite[eq. (2.3.6/9)]{prudnikov_book_vol1}, $\mathcal{T}_{3,i}(a_i,b_i,c_i)$ can be solved in closed form as
\begin{equation}\label{eq:capacity_exact_solution}
\begin{split}
\mathcal{T}_{3,i}(a_i,b_i,c_i)=&\sum_{j=1}^{i+1}\sum_{h=1}^{b_j} \Xi_{j,h}\Gamma(D+1) a_j^{D+1-h}  \\ & \times \Psi \left[ D+1,D+2-h, \left(\sum_{j=1}^ic_j\right)\right].
\end{split}
\end{equation}
\paragraph{Upper Bound}For obtaining a bound on the capacity performance, the Jensen's inequality is applied and thus the following upper bound on the capacity is obtained
\begin{equation}\label{eq:ergodic_capacity_bound}
C_{\gamma_{\rm tot,k}}\leq\frac1{NH} \log_2\left(1+\int_0^\infty \gamma f_{\gamma_{\rm tot,k}}(\gamma)d\gamma\right).
\end{equation}
Substituting the PDF expression provided in \ref{sub:eNB_UEs} (for the conventional cellular case) or \ref{par:total_hybrid_statistics} (for the hybrid Cellular/WLAN case) in \eqref{eq:ergodic_capacity_bound}, integrals of the form given in \eqref{eq:SINR_integrals_definition} appear. By following the analysis presented in \ref{subs:ASINR_analysis}, a closed form solution for this integral can be derived. Therefore, using these solutions, the ergodic capacity of the conventional cellular scenario can be upper bounded as
\begin{equation}\label{eq:enb_UE_capacity_final}
\begin{split}
C_{\gamma_{\textrm {c,k}} }(\gamma)&\leq\log_2\left[ 1+\frac{\overline{\gamma}_{{eu_{\textit{\scriptsize n,k}}}}}{\overline{\gamma}_{I_{{eu,k}}}^2} \Psi\left( 2,3-|\mathcal{I}_{UE,k}^C|,\frac1{\overline{\gamma}_{I_{{eu,k}}}}\right) \right.\\ & \left. +|\mathcal{I}_{UE,k}^C| \frac{\overline{\gamma}_{{eu_{\textit{\scriptsize n,k}}}}}{\overline{\gamma}_{I_{{eu,k}}}} \Psi\left( 2,2-|\mathcal{I}_{UE,k}^C|,\frac1{\overline{\gamma}_{I_{{eu,k}}}}\right)\right].
\end{split}
\end{equation}

It should be noted that the previously presented approach can be also directly applied to the hybrid Cellular/WLAN scenario for obtaining the capacity. However, the extracted expression for the capacity is not included here due to space limitations.
\begin{figure*}[]
\begin{equation}\label{eq:MU_SINR}
\gamma_{\textrm{k}}=
\begin{rcases}
  &\frac{\gamma_{eu_{\textit{\scriptsize n,k}}}}{1+\sum\limits_{i=1}^{|\mathcal{I}_{AP,k}^{C}|} \gamma_{I_{ea,i}} + \sum\limits_{i=1}^{|\mathcal{I}_{UE,k}^{C}|}\gamma_{I_{eu,i}}}, \hspace{5cm} &\text{if UE served by the $n$th eNB \ } \\
  &min \left\{\frac{\gamma_{_{ea_{\textit{\scriptsize n,m}}}}}{1+\sum\limits_{i=1}^{|\mathcal{I}_{AP,k}^{C}|} \gamma_{I_{ea,i}} + \sum\limits_{i=1}^{|\mathcal{I}_{UE,k}^{C}|}\gamma_{I_{eu,i}}}, \frac{\gamma_{_{au_{\textit{\scriptsize m,k}}}}}{1+\sum\limits_{i=1}^{|\mathcal{I}_{UE}^{W}|} \gamma_{I_{au,i}}}    \right\}, \hspace{0.5cm} &\text{if UE served by an WLAN AP}
\end{rcases}
\end{equation}
\begin{center}
\line(1,0){500}
\end{center}
\end{figure*}

\subsubsection{Outage Probability (OP)}
The OP is defined as the probability that the SINR falls below a predetermined threshold $\gamma_\textrm {th} $ and is given by $P_{\gamma_{ \textrm {tot,k}}}=F_{\gamma_{ \textrm {tot,k}}}(\gamma_{\textrm {th}})$. Based on this definition, the OP can be evaluated by using the CDF expression provided in \ref{sub:eNB_UEs} (for the conventional cellular case) or \ref{par:total_hybrid_statistics} (for the hybrid Cellular/WLAN case).

\vspace{0.0cm}

\section{Multi-Cellular Performance Optimization}
Considering the importance of mitigating the co-channel interference in MU multi-cellular networks, in this Section, the proposed approach is extended to a MU multi-cellular communication scenario, aiming to optimize the system performance by minimizing the intercell interference via the utilization of WLAN APs, which are distributed within the cells. More specifically, we consider a network setup with $N,M$ intercell interfering UEs and APs, respectively. Furthermore, assuming that each UE is connected to a specific eNB (or a WLAN AP), the SINR of the $k$th UE is given by \eqref{eq:MU_SINR} at the top of this page. Then the overall average instantaneous SINR is expressed as \vspace{-0.0cm}
\begin{equation}\label{eq:MU_SINR_total}
SINR_{tot} = \frac{\sum \limits_{k=1}^N  \gamma_{\textrm{k}}}{N}.
\end{equation}

The optimum network setup in terms of the SINR maximization (or equivalently the sum-rate maximization) is determined by the sets $\left\{ \mathcal{I}_{UE,k}^{C}, \mathcal{I}_{AP,k}^{C}, \mathcal{I}_{UE}^{W}\right\}$, which minimize the total intercell interference. Therefore, the optimization problem may be expressed as
\begin{equation}\label{eq:MU_SINR_opt}
\underset{\left\{ \mathcal{I}_{UE,k}^{C}, \mathcal{I}_{AP,k}^{C}, \mathcal{I}_{UE,k}^{W}\right\}, \forall k}{SINR_{opt} \text{ = arg max}\left\{ SINR_{tot} \right\}}.
\end{equation}

The optimal solution to the above problem can be found via exhaustive search, which however can be very time consuming, since the number of searches equals to $(N+M)!$ (e.g., $3,628,800$ for 4 eNBs, 4 UEs and 6 APs). To this end, we propose a suboptimal greedy algorithm which considerably reduces the number of searches among all possible network combinations. The mode of operation of this algorithm, which is presented in detail in Table IV, is summarized as follows. For a network setup with $N$ co-channel interfering UEs and $M$ APs, the algorithm initially considers only the existing cellular topology, ignoring the existence of the WLAN APs. Then starting by randomly selecting a UE that is served by an eNB, the serving eNB is replaced by a WLAN AP connected to that eNB and the total SINR is calculated, assuming no other changes to the rest of the UEs connections. If the resulting total SINR increases, then this AP is considered as the serving AP for the first UE, otherwise UE continues to be served by its eNB. This process is repeated until all the WLAN AP are tested as the candidate AP for the first UE. The AP that results in the highest SINR is considered the serving AP. After finishing with the first UE, the algorithm repeats the same process for the remaining UEs. Obviously, the total number of searches is given by $N*M$.
\begin{table}[]
\renewcommand{\arraystretch}{1.1}
\caption {GREEDY ALGORITHM for OVERALL SINR MAXIMIZATION (MU Multi-Cell Network)}
\label{Tab:1} \centering
\begin{tabular}{l  }
  \hline
 {\bf Input:} \\
 \text{$\bullet$ The number of co-channel interfering UEs, $N$,} \\
 \text{\ \ \ the number of WLAN APs, $M$. } \\
 \text{$\bullet$ The channel gains between the eNBs and the UEs.}   \\
 \text{$\bullet$ The channel gains between the eNBs and the APs. }\\
 \text{$\bullet$ The channel gains between the APs and the UEs. } \\
 \\
 \underline{\textbf{Initialization stage}}\\
 \text{Determine the initial cellular topology,}\\
  \text{based on the minimum distance criterion.}\\
   \text{Calculate the corresponding total SINR.}\\

 \\

\underline{\textbf{Greedy stage}}\\
{\bf for} $j:=1$ to $N$ \\
\text{Considering the topology $\mathcal{S}(1)=[t(1) \ t(2) \ ... \ t(j-1) \ t(j)]$} \\
\text{ \ \ \ \  \ }{\bf for} $a:=N+1$ to $N+1+M$ \\
\text{ \ \ \ \  \ }\text{Calculate the overall SINR for the topology} \\
\text{ \ \ \ \  \ }\text{$\mathcal{S}(a)=[t(a) \ t(2) \ ... \ t(j-1) \ t(j)]$} \\
\text{ \ \ \ \  \ }\text{\bf{If}} \text{the resulting SINR with topology $\mathcal{S}(a)$} increases, \\
\text{ \ \ \ \  \ }\text{then the $a$th WLAN AP becomes the serving AP for the $j$th UE} \\
\text{ \ \ \ \  \ }\text{\bf{else}} \text{the topology remains as it is.} \\
\text{ \ \ \ \  \ }{\bf end for}\\
{\bf end for}\\
\\
\hline
\end{tabular}
\end{table}

{\textbf{Example:}}
Consider a multi-cellular network with $N=3$ co-channel interferers, and $M=2$ WLAN APs and assume that the initial cellular topology is denoted by the vector $\mathcal{S}=[2 \ 3 \ 1]$, where the $k$th element of the vector denotes the AP ($(1,..., N)$ for the eNB and $(N+1,..., N+1+M)$ for the rest of WLAN APs) that the $k$th UE is connected to. Here the vector $\mathcal{S}=[2 \ 3 \ 1]$ denotes that the $1$st UE is connected to the $2$nd eNB, the $2$nd UE is connected to the $3$rd eNB and the $3$rd UE to the $1$st eNB. Then the algorithm starts testing the WLAN APs as candidate serving AP for the UEs. For the first UE the candidate new network topology will be $\mathcal{S}=[4 \ 3 \ 1]$ and the corresponding total SINR will be calculated. Assume that the resulting SINR does not increase, so the $2$nd eNB continues to serve the $1$st UE. The algorithm continues testing the $2$nd WLAN AP and the candidate new network topology will be $\mathcal{S}=[5 \ 3 \ 1]$. Assume that the resulting SINR increases, so the $2$nd AP will be the new serving AP for the $1$st UE. Then considering the new topology, the algorithm will continue the same process for the $2$nd UE and will test all the WLAN APs as candidate serving APs.

\section{A Novel Scheme for Frequency Reuse Equal to One within a Macrocell}
In this Section, based on the general multi-cellular hybrid Cellular/WLAN concept, we present a novel scheme that could be used for achieving frequency reuse equal to one within a single macro-cell, under specific performance criteria and constraints that guarantee the overall cell's or the individual UE's QoS requirements. Such a scheme could be also used for cognitive radio systems, where the secondary UEs could utilize the same frequencies as the primary UEs, provided that the each UE satisfies the desired QoS requirements.
\begin{figure*}[t!]
\centering
\includegraphics[keepaspectratio,width=10.8cm, trim=2cm 0cm 2cm 2cm]{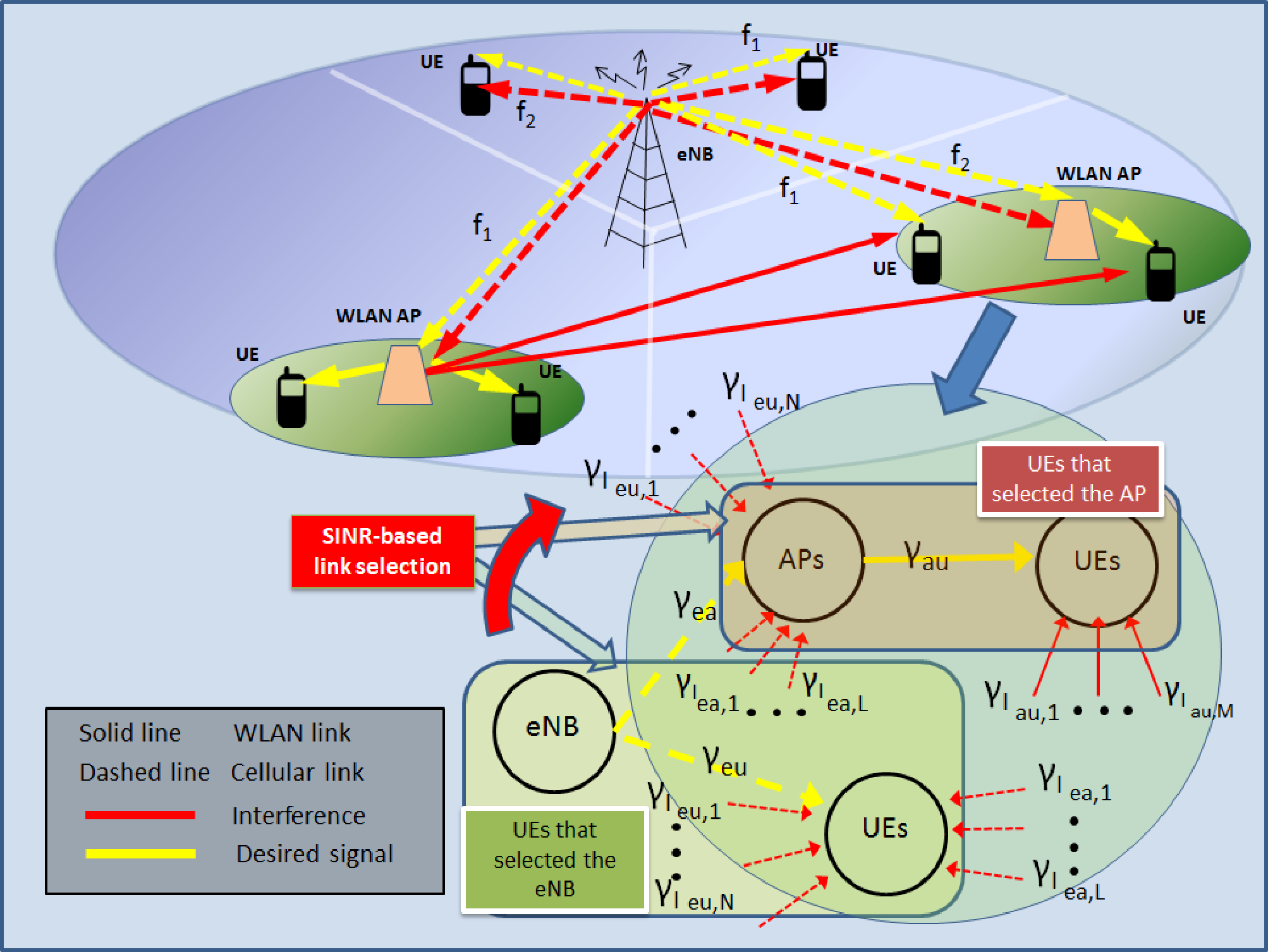}
\caption{A single macrocell network with frequency reuse one. } \label{Fig2}
\end{figure*}

Let us consider a single macrocell network, where $N$ UEs and $M$ APs may reuse the same frequency. Assuming that we have determined the serving point (eNB or WLAN AP) that each UE will be connected to, the SINR of the $k$th UE is given by \eqref{eq:MU_SINR}, where now only one eNB exists, i.e., $n=1$, and the overall average cell's instantaneous SINR is expressed as in \eqref{eq:MU_SINR_total}. Under these assumptions, two optimization problems may be considered as follows.

\underline{{Overall cell SINR Maximization}}

\begin{table}[]
\renewcommand{\arraystretch}{1.1}
\caption {GREEDY ALGORITHM for OVERALL SINR MAXIMIZATION (MU Single-Cell Network)}
\label{Tab:1} \centering
\begin{tabular}{l  }
  \hline
 {\bf Input:} \\
 \text{$\bullet$ The number of UEs, $N$, the number of WLAN APs, $M$. } \\
 \text{$\bullet$ The number of UEs, $Q$, for the optimal topology} \\
 \text{\ \ \ in the initialization stage. } \\
 \text{$\bullet$ The channel gains between the eNB and the UEs.}   \\
 \text{$\bullet$ The channel gains between the eNB and the APs. }\\
 \text{$\bullet$ The channel gains between the APs and the UEs. } \\
 \\
 \underline{\textbf{Initialization stage}}\\
 \text{Considering $Q$ UEs, find the optimal topology} \\
 \text{that maximizes the overall SINR} \\
 \text{Save the optimal topology in the vector $\mathcal{S}=[t(1) \ t(2) \ ... \ t(Q)]$, }\\
 \text{where $t(\cdot)=1$ if the AP is the eNB  }\\
  \text{and $t(\cdot)=M+1$ if the AP is the $M$th WLAN AP.}\\
 \\

\underline{\textbf{Greedy stage}}\\
{\bf for} $j:=Q+1$ to $N$ \\
\text{Considering the topology $\mathcal{S}(1)=[t(1) \ t(2) \ ... \ t(j-1) \ t(j)]$} \\
\text{ \ \ \ \  \ }{\bf for} $a:=1$ to $M+1$ \\
\text{ \ \ \ \  \ }\text{Calculate the overall SINR for the topology} \\
\text{ \ \ \ \  \ }\text{$\mathcal{S}(a)=[t(1) \ t(2) \ ... \ t(j-1) \ t(j) \ t(a)]$} \\
\text{ \ \ \ \  \ }{\bf end for}\\
\text{Among all $\mathcal{S}(a)=[t(1) \ t(2) \ ... \ t(j-1) \ t(j) \ t(a)]$} \\
\text{Find the topology that maximizes the total mean SINR} \\
\text{Set $\mathcal{S}(j)=[t(1) \ t(2) \ ... \ t(j-1) \ t(j) \ t(a)]$ } \\
{\bf end for}\\
\\
\hline
\end{tabular}
\end{table}

\begin{table}[]
\renewcommand{\arraystretch}{1.1}
\caption {GREEDY ALGORITHM for OVERALL SINR MAXIMIZATION with FAIRNESS (MU Single-Cell Network)}
\label{Tab:1} \centering
\begin{tabular}{l  }
  \hline
 {\bf Input:} \\
 \text{$\bullet$ The number of UEs, $N$, the number of WLAN APs, $M$. } \\
 \text{$\bullet$ The number of UEs, $Q$, for the optimal topology} \\
 \text{\ \ \ in the initialization stage. } \\
 \text{$\bullet$ The channel gains between the eNB and the UEs.}   \\
 \text{$\bullet$ The channel gains between the eNB and the APs. }\\
 \text{$\bullet$ The channel gains between the APs and the UEs. } \\
 \\
 \underline{\textbf{Initialization stage}}\\
 \text{Considering $Q$ UEs, find the optimal topology} \\
 \text{that maximizes the minimum individual SINR among all UEs} \\
 \text{Save the optimal topology in the vector $\mathcal{S}=[t(1) \ t(2) \ ... \ t(Q)]$, }\\
 \text{where $t(\cdot)=1$ if the AP is the eNB  }\\
  \text{and $t(\cdot)=M+1$ if the AP is the $M$th WLAN AP.}\\
 \\

\underline{\textbf{Greedy stage}}\\
{\bf for} $j:=Q+1$ to $N$ \\
\text{Considering the topology $\mathcal{S}(1)=[t(1) \ t(2) \ ... \ t(j-1) \ t(j)]$} \\
\text{ \ \ \ \  \ }{\bf for} $a:=1$ to $M+1$ \\
\text{ \ \ \ \  \ }\text{Calculate the individual SINR of each UE for the topology} \\
\text{ \ \ \ \  \ }\text{$\mathcal{S}(a)=[t(1) \ t(2) \ ... \ t(j-1) \ t(j) \ t(a)]$} \\
\text{ \ \ \ \  \ }{\bf end for}\\
\text{Among the $\mathcal{S}(a)=[t(1) \ t(2) \ ... \ t(j-1) \ t(j) \ t(a)]$} \\
\text{Find the topology that maximizes the minimum individuial SINR} \\
\text{Set $\mathcal{S}(j)=[t(1) \ t(2) \ ... \ t(j-1) \ t(j) \ t(a)]$ } \\
{\bf end for}\\
\\
\hline
\end{tabular}
\end{table}
The aim of this problem is to maximize the overall cell SINR and hence the sum-rate, independently from  the individual UE SINR requirements. The corresponding optimization problem is expressed as in \eqref{eq:MU_SINR_opt} but assuming a single macro-cell, i.e.,
\begin{equation}\label{eq:MU_SINR_opt_single}
\underset{\left\{ \mathcal{I}_{UE,k}^{C}, \mathcal{I}_{AP,k}^{C}, \mathcal{I}_{UE,k}^{W}\right\}, \forall k}{SINR_{opt} \text{ = arg max}\left\{ SINR_{tot}|_{n=1} \right\}}
\end{equation}

\underline{{Overall cell SINR Maximization with Fairness}}
Maximizing the overall cell SINR without considering the individual SINR requirements of each UE may result with high probability that each time only a limited number of UEs are benefited, while other UEs may be unable to meet their QoS requirements. In order to avoid this situation and attain the fairness among the UEs (obviously at the cost of the overall cell SINR), we also consider the following optimization problem

\begin{gather}\label{eq:MU_SINR_opt_fair}
\underset{\left\{ \mathcal{I}_{UE,k}^{C}, \mathcal{I}_{AP,k}^{C}, \mathcal{I}_{UE,k}^{W}\right\}, \forall k}{SINR_{opt} \text{ = arg max}\left\{ min(SINR_{k}) \right\}}.
\end{gather}

The optimal solution to these two optimization problems is found via exhaustive search, which however can be very time consuming for a relative high number of APs and UEs since the number of searches equals to $E=(M+1)^N$ (e.g., $279,936$ for $N=7$ and $M=5$). To this end, similar to the multi-cell case, we propose a suboptimal greedy algorithm which considerably reduces the number of searches among all possible network combinations. The mode of operation of this algorithm, which is presented in detail in Table V, is summarized as follows. For a network setup with $N$ UEs and $M$ APs, the algorithm initially considers only $Q$ of the total UEs and finds among all the $(M+1)^Q$ the optimal network topology that maximizes the overall cell SINR. Then continuing to the $Q+1$ UE, the algorithm looks for the AP that this UE will be connected to, searching among the $M$ APs, in order to maximize the overall SINR given that the previous $Q$ UEs cannot change their AP. Similarly, considering that the previous $Q+1$ UEs cannot change their APs, the algorithm continues to the $Q+2$ UE and searches for that AP that the UE will be connected to, in order to maximize the overall SINR. In this way the total number of searches is given by $G=(M+1)^Q + (N-Q)*M$.

\underline{\textbf{Example:}}
Consider a macro-cell with an eNB, $N=5$ UEs and $M=3$ WLAN APs and assume that the optimal network topology, which is found via exhaustive search for the first 2 UEs, is denoted by the vector $\mathcal{S}=[2 \ 1]$, where the $k$th element of the vector denotes the serving point ($1$ for the eNB and $(2,..., M+1)$ for the rest of WLAN APs) that the $k$th UE is connected to. Here the vector $\mathcal{S}=[2 \ 1]$ denotes that the $1$st UE is connected to the $2$nd WLAN AP and the $2$nd UEs is connected to the eNB. Then the algorithm continues to the $3$rd UE and searches among the $M=3$ APs, for the AP that will provide the maximum overall SINR, i.e., the algorithm will calculate the overall SINR for the following topologies: $\mathcal{S}=[2 \ 1 \ 1]$, $\mathcal{S}=[2 \ 1 \ 2]$, $\mathcal{S}=[2 \ 1 \ 3]$, $\mathcal{S}=[2 \ 1 \ 4]$ and will select the topology that maximizes the total SINR. Assuming that the algorithm selects the topology $\mathcal{S}=[2 \ 1 \ 3]$, it will continue to consider the $4$th UE and will search among the topologies $\mathcal{S}=[2 \ 1 \ 3 \ 1]$, $\mathcal{S}=[2 \ 1 \ 3 \ 2]$, $\mathcal{S}=[2 \ 1 \ 3 \ 3]$ and $\mathcal{S}=[2 \ 1 \ 3 \ 4]$. The algorithm continues until all UEs are connected to a service point.

Considering the maximization of the SINR with fairness, the algorithm's mode of operation is very similar (see Table VI), with the difference that in the initialization stage the algorithm finds via exhaustive search the topology that maximizes the minimum SINR among the UEs.

\section{Numerical Results and Discussion}
In this section, we present and discuss several numerical results, which demonstrate the performance of the proposed architecture compared to the conventional cellular system, which serves as a benchmark. Moreover, numerical simulations are performed to validate the mathematical analysis.
\begin{figure}[b!]
\centering
\includegraphics[keepaspectratio,width=2.35in, trim=2cm 0cm 1cm 0.0cm]{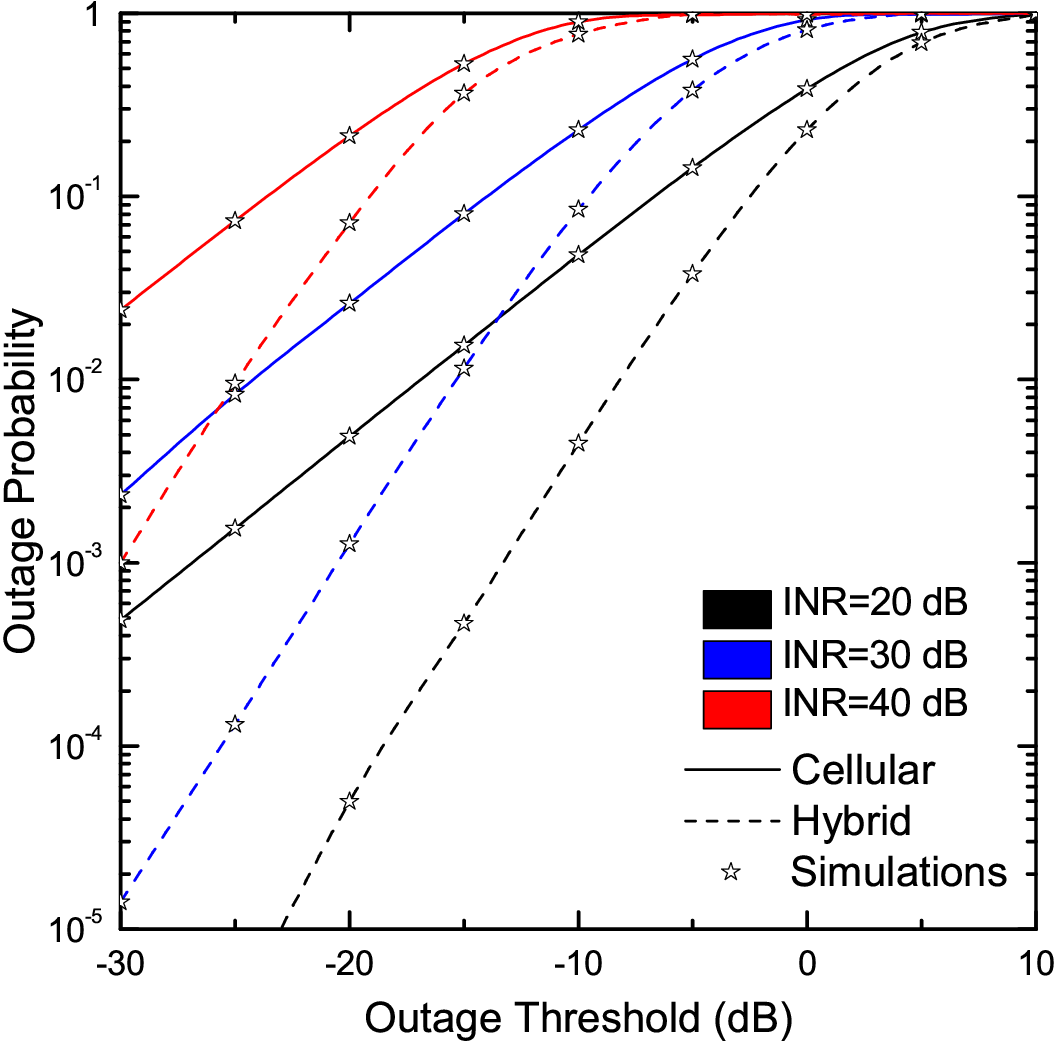}
\caption{OP comparisons as a function of the outage threshold for various values of the average INR. } \label{Fig2}
\end{figure}

\begin{figure}[t!]
\includegraphics[keepaspectratio,width=2.4in, trim=-2cm 0cm 4.5cm 0.0cm]{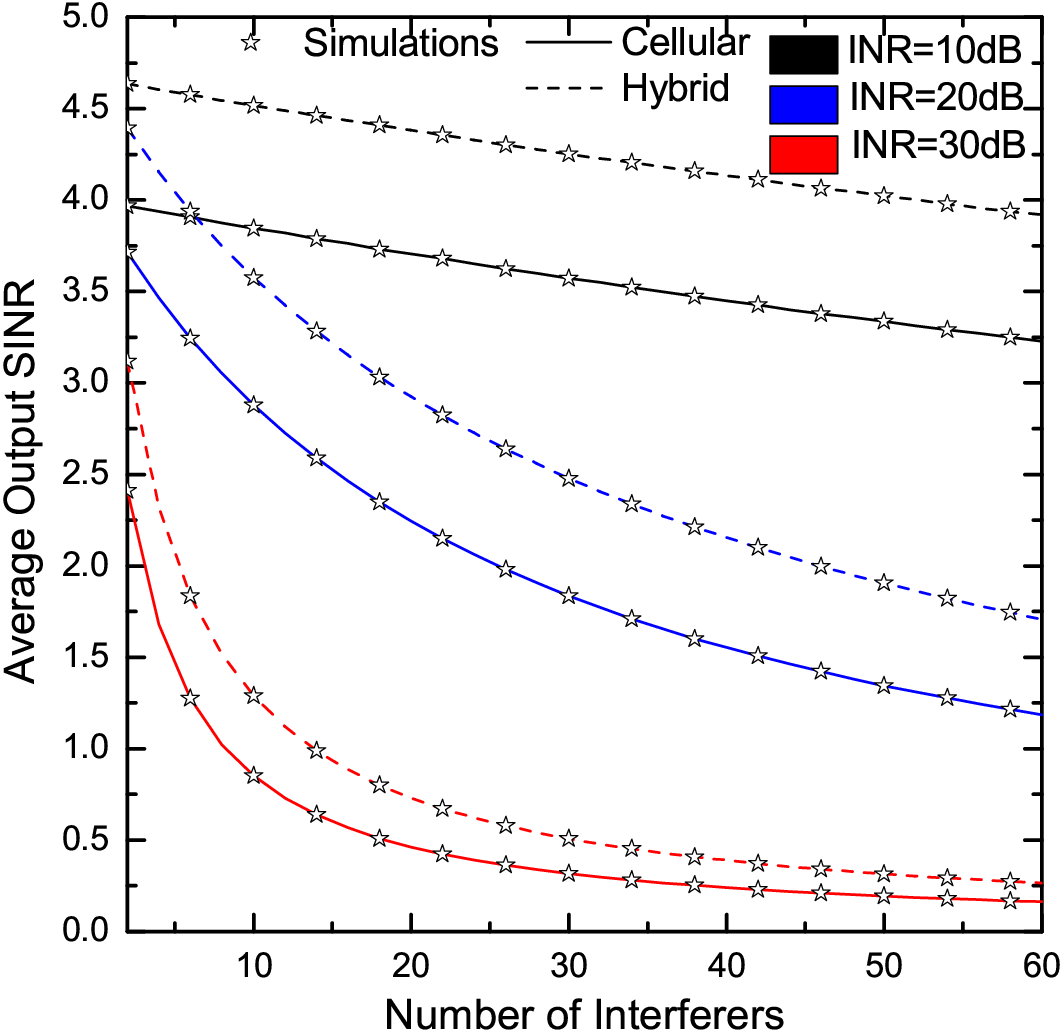}
\caption{Average output SINR comparisons as a function of the outage number of interferers for various values of the average INR. } \label{Fig3}
\end{figure}

In Fig.~\ref{Fig2}, a comparison of the OPs between the conventional scheme and the proposed hybrid approach are plotted as a function of the outage threshold $\gamma_{\textrm th}$. In this figure, for the conventional cellular scheme we have assumed $|\mathcal{I}_{UE,k}^C|=24$, for the hybrid $|\mathcal{I}_{UE,k}^{C}|=|\mathcal{I}_{AP,k}^{C}|=|\mathcal{I}_{UE}^{W}|=12$, while in all cases the average transmit power is equal to $40$dB. It is depicted that the hybrid scheme provides a significant performance improvement. Additionally the performance gap between these two communication approaches increases as the average INR decreases.

In Fig.~\ref{Fig3}, the average output SINR of both schemes is plotted as a function of the total number of interferers, for various values of the average INR. In this figure, it is also depicted that the proposed scheme significantly improves the performance in all cases, compared to the conventional one, while the performance gap decreases as the INR increases. Therefore, it becomes evident that the total interference within a cell can be considerably reduced if wireless WLAN APs are deployed within the cell and utilized for sharing their broadband connection with adjacent users. Furthermore, these results show that by employing low cost WLAN APs, which require no additional infrastructure (in contrast to small-cell deployments), the average SINR can be increased compared with a conventional cellular network, where all users are served directly by the eNB.
\begin{figure}[t!]
\centering\vspace*{2cm}
\includegraphics[keepaspectratio,width=2.9in]{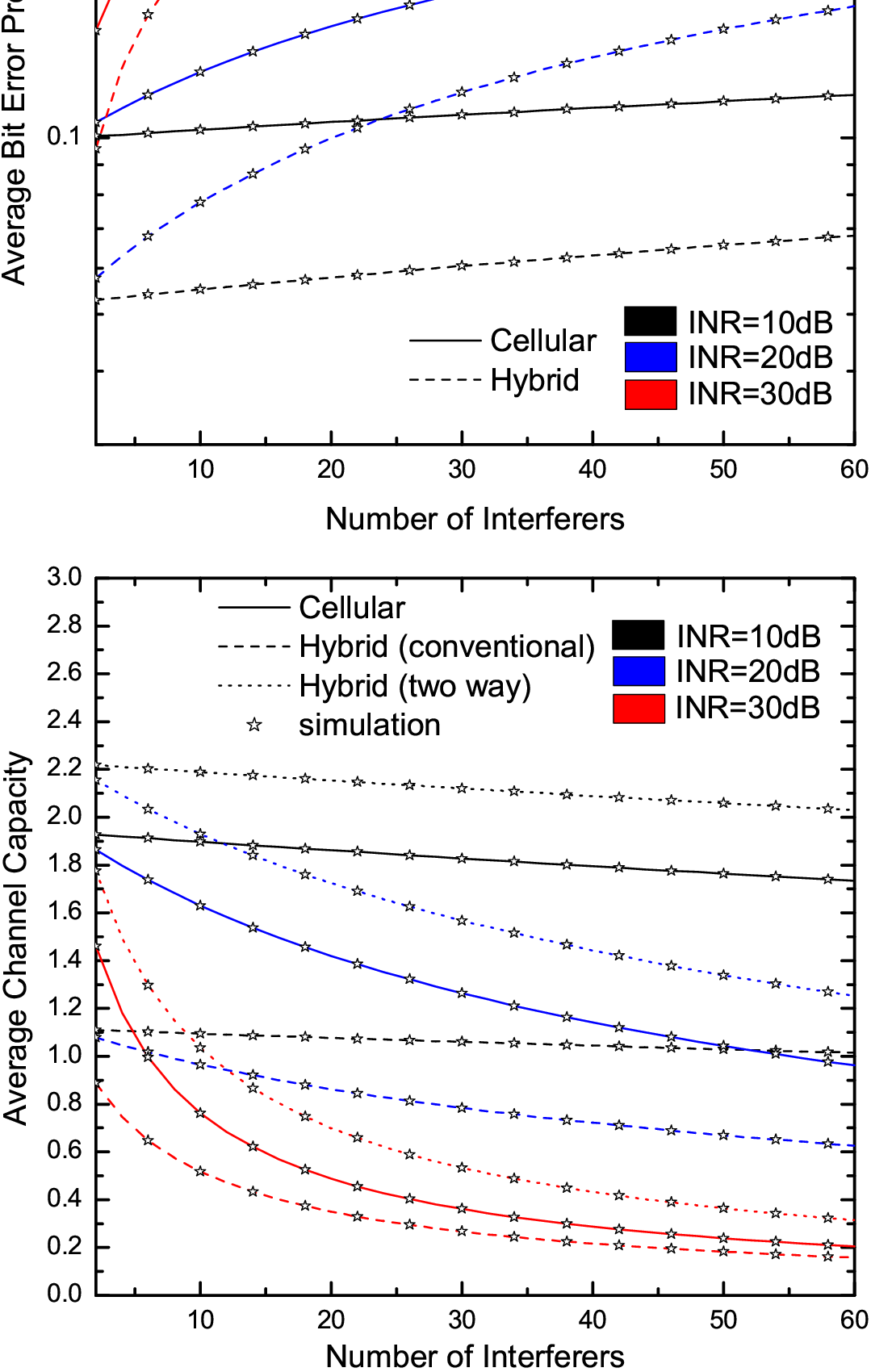}
\caption{The ABEP and channel capacity of the hybrid as well as the cellular communications systems as a function of the number of interferers.} \label{Fig:Fig4}
\end{figure}

In Fig.~\ref{Fig:Fig4}, considering average transmit power equal to $40$dB, the average ABEP as well as channel capacity of both schemes are plotted as a function of the number of interferers also for various values of the average INR. In this figure, it is clearly depicted that the ABEP performance improves by employing the hybrid scheme. However, since the relaying scheme under consideration is one way, its information theoretic capacity is divided by a factor of two, resulting in an decrease in the overall hybrid scheme capacity. However, in many works it has been shown that by adopting a two-way relay channel, it is possible to remove the $1/2$ loss rate loss factor in
the capacity, e.g., \cite{4217667,6034723}. Following such an approach in our case, the hybrid system capacity is considerably improved, thus providing the best performance in all cases. Finally, for comparison purposes, computer simulation results are also included in Figs.~\ref{Fig2}-\ref{Fig:Fig4}, verifying the validity of the proposed theoretical approach.

\begin{figure}[]
\centering
\includegraphics[keepaspectratio,width=8.5cm, trim=2cm 0cm 2cm 2cm]{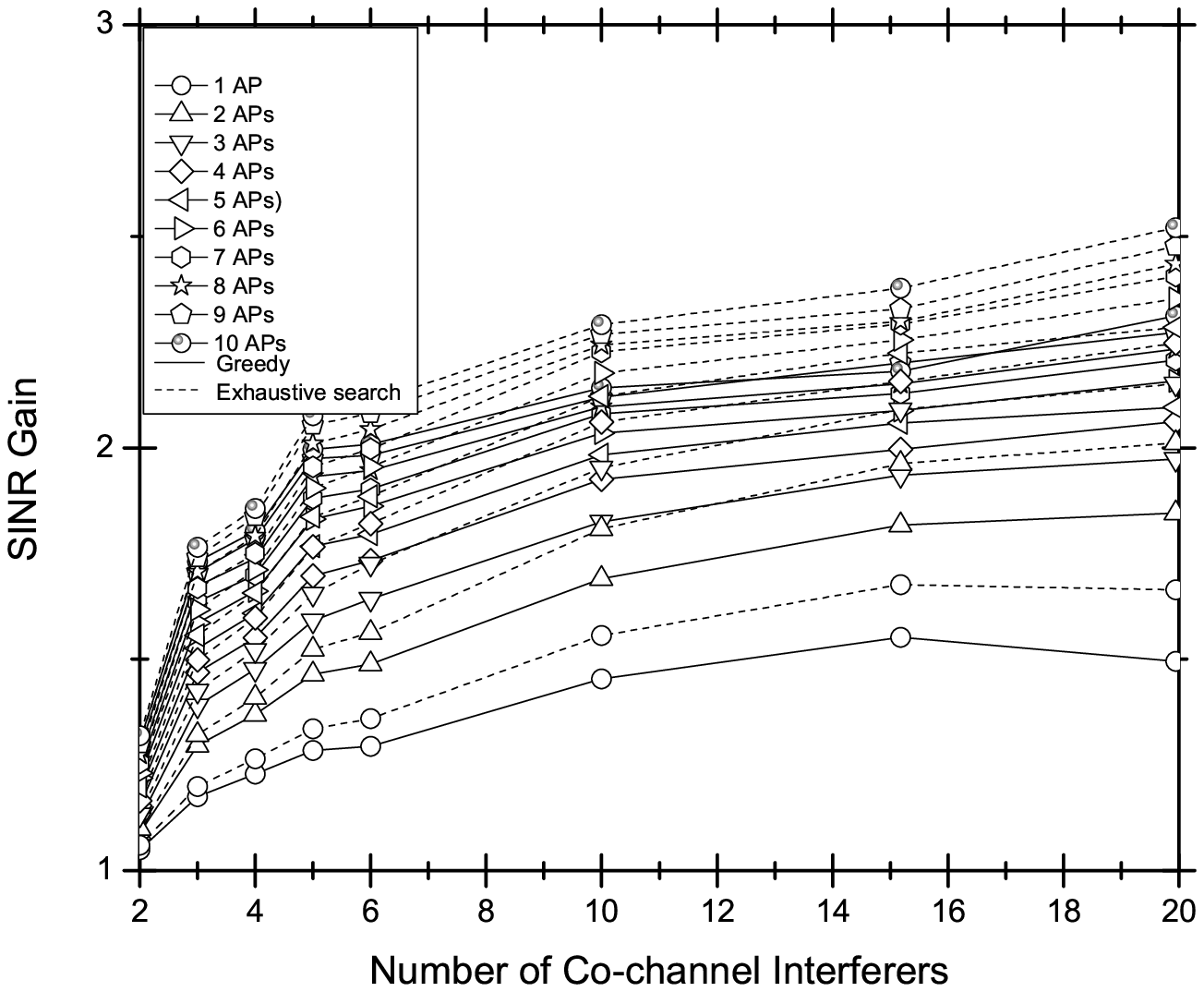}
\caption{The SINR gain of the proposed Cellular/WLAN scheme over the cellular (MU multi-cell scenario).} \label{Fig:Fig_MU_MC}
\centering
\includegraphics[keepaspectratio,width=8.5cm, trim=2cm 0.0cm 2cm 1cm]{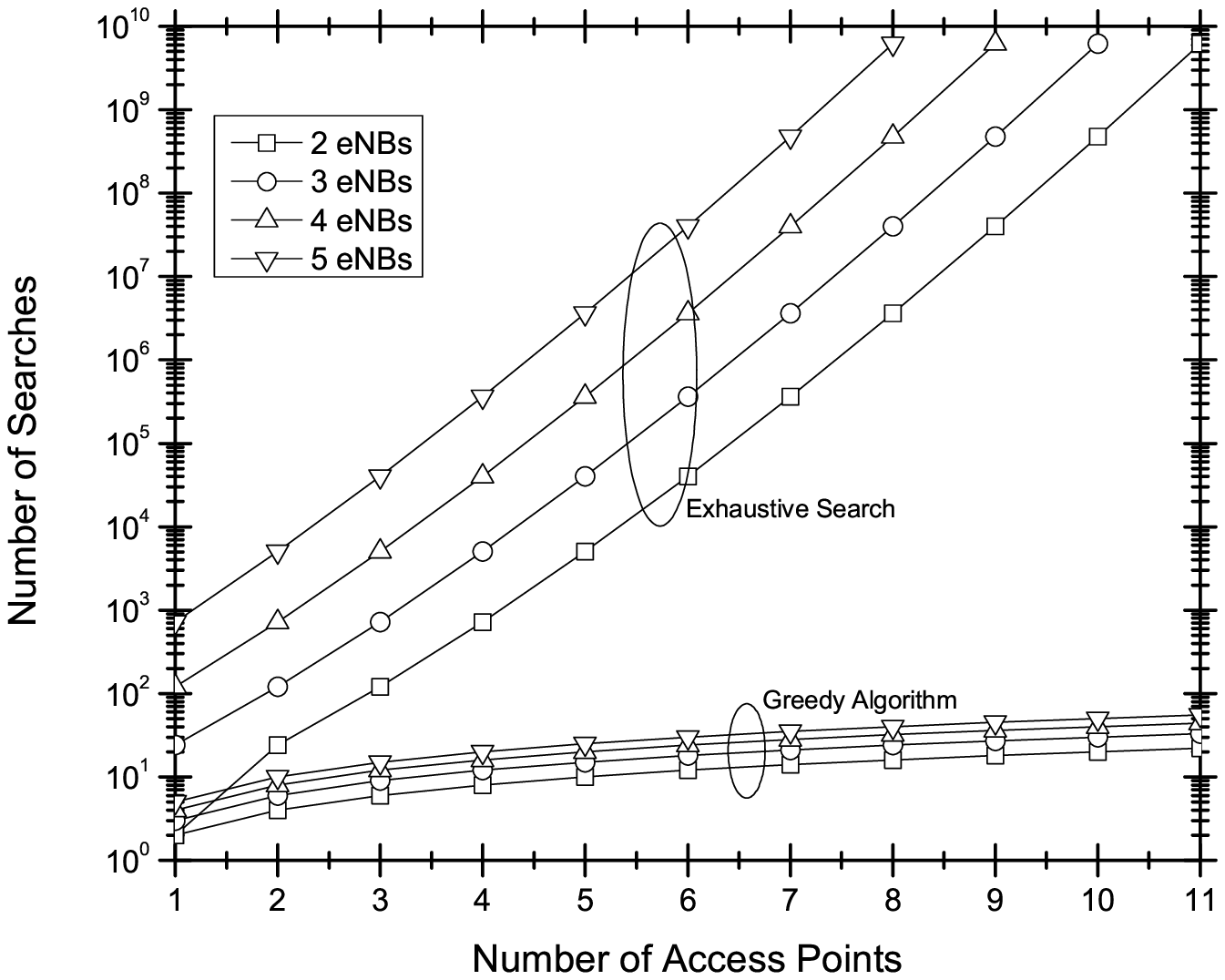}
\caption{The number of searches of the exhaustive and the greedy solution.} \label{Fig:Fig_MU_iter_MC}
\end{figure}

The theoretical results provide insight into the performance of a single UE given that the network topology has been determined, i.e., each UE has selected the AP to be connected to. Nevertheless, determining the optimal network topology is a challenging task. As mentioned in Section V, the straightforward method is to exhaustively search among all the possible network topologies and select the one that maximizes the performance metric. However, the number of searches may be prohibitive even for topologies with less than 4 UEs and 4 APs. The simulation setup for the MU case considers the downlink of a cellular network with cell radius equal to $500$m, where the UEs and APs are uniformly distributed within the cell. The frequency of the cellular interferers is $800$MHz, while the frequency of the WLAN interferers is $2.4$GHz. The maximum eNB transmission power is $10$W, the maximum power of a WLAN AP is $0.1$W and the noise power is $10^{-10}$ W/Hz. The channel gains are exponentially distributed with average value of the received signal power calculated by the Friis propagation model. For the averaging of the results, $2000$ different random topologies are used.
Considering the multi-cell MU case, the SINR gain of the hybrid Cellular/WLAN scheme over the baseline conventional cellular is given in Fig.~\ref{Fig:Fig_MU_MC} as a function of the number of co-channel interference. It is noted that with the number of interferers we denote the number of UEs using the same frequency. In the conventional cellular scheme, the UEs connect to the eNB that maximizes the overall SINR. We observe the gain increases as the number of interferers increases until it reaches a ceiling. The greedy solution is close to the optimum one obtained via exhaustive search, especially for a small number of APs. The number of searches required by the optimum solution and the greedy one is depicted in Fig.~\ref{Fig:Fig_MU_iter_MC}.

\begin{figure}[]
\centering
\includegraphics[keepaspectratio,width=9.5cm, trim=2cm 0cm 1cm 2cm]{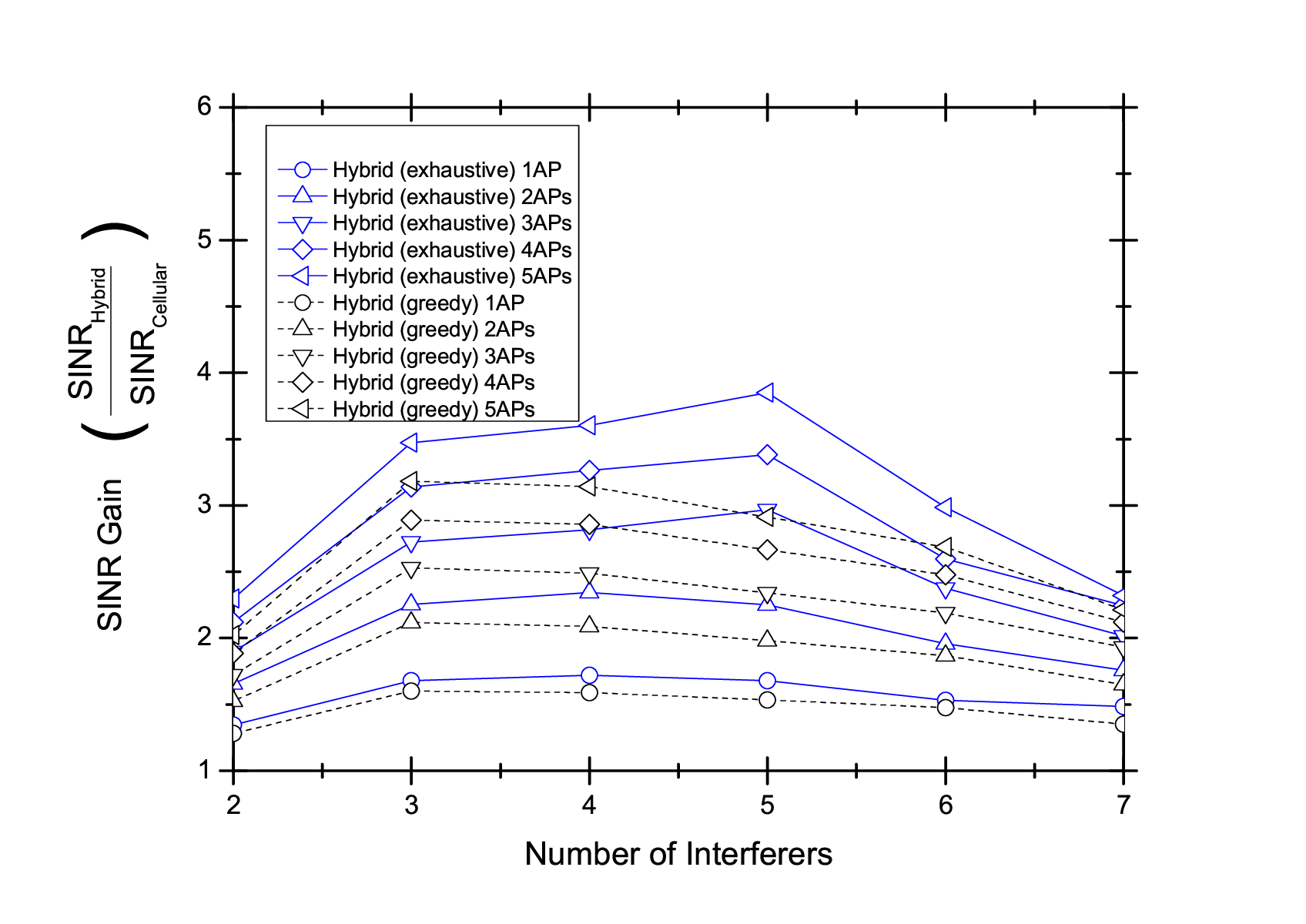}
\caption{The SINR gain of the proposed Cellular/WLAN scheme over the cellular (Single cell scenario).} \label{Fig:Fig_MU}
\centering
\includegraphics[keepaspectratio,width=9.5cm, trim=2cm 0.0cm 1cm 0.0cm]{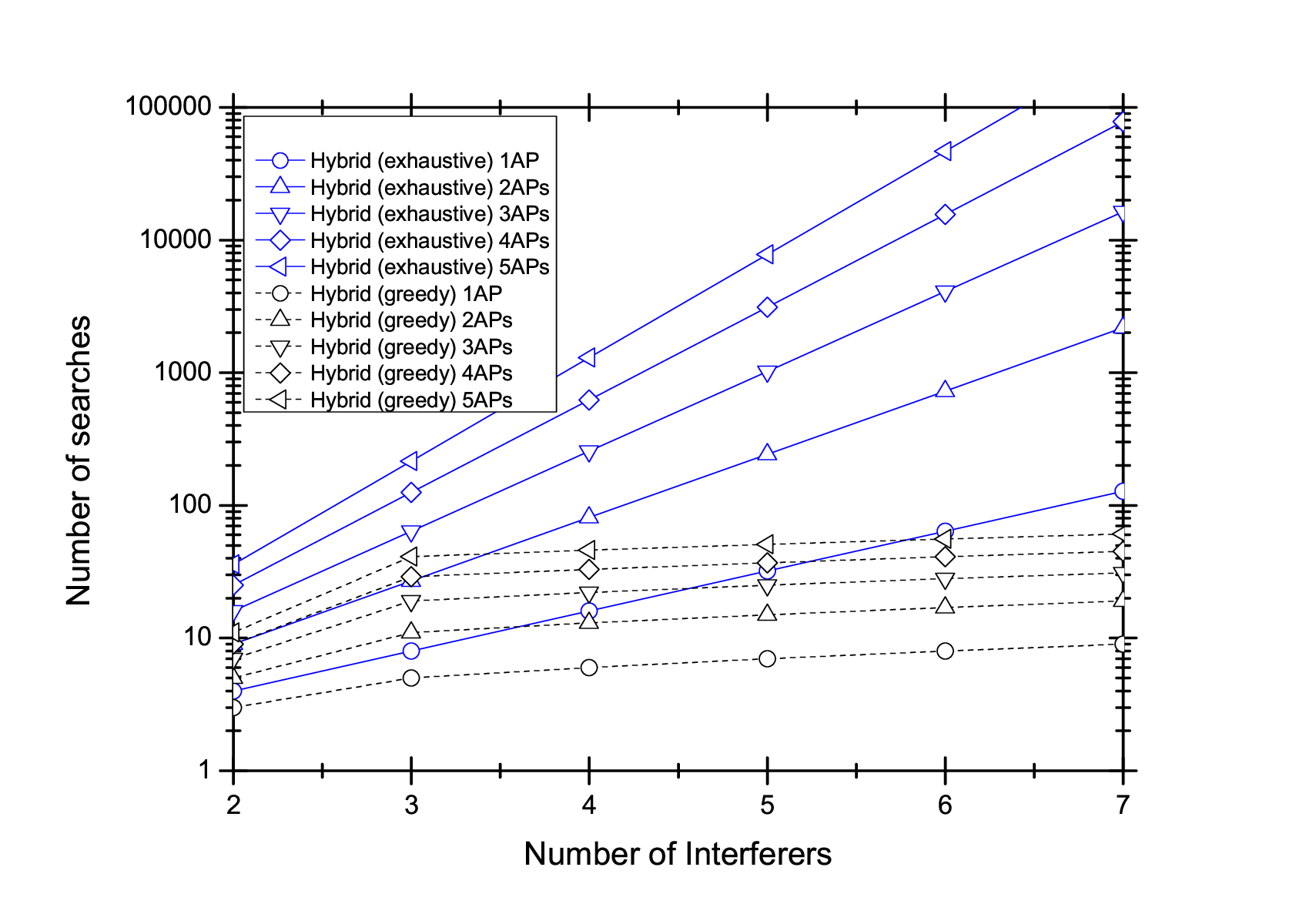}
\caption{The number of searches of the exhaustive and the greedy solution.} \label{Fig:Fig_MU_iter}
\end{figure}

\begin{figure}[t!]
\includegraphics[keepaspectratio,width=9.0cm, trim=2cm 0.0cm 2cm 2cm]{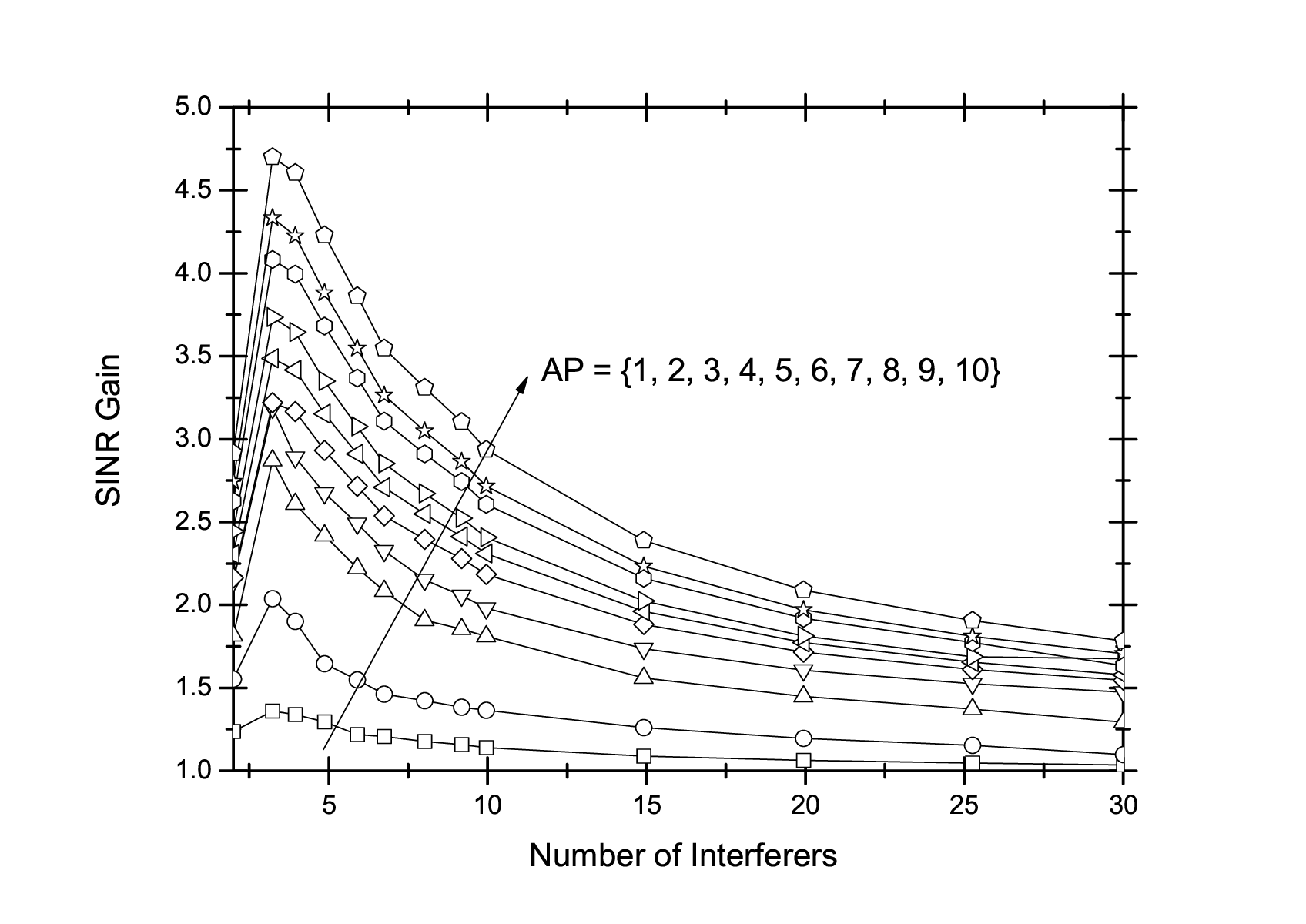}
\caption{The SINR gain of the proposed Cellular/WLAN scheme over the cellular ((Single cell scenario, Greedy algorithm).} \label{Fig:Fig_MU_greedy}
\end{figure}

\begin{figure}[t!]
\includegraphics[keepaspectratio,width=9.0cm, trim=2cm 0.0cm 2cm 2cm]{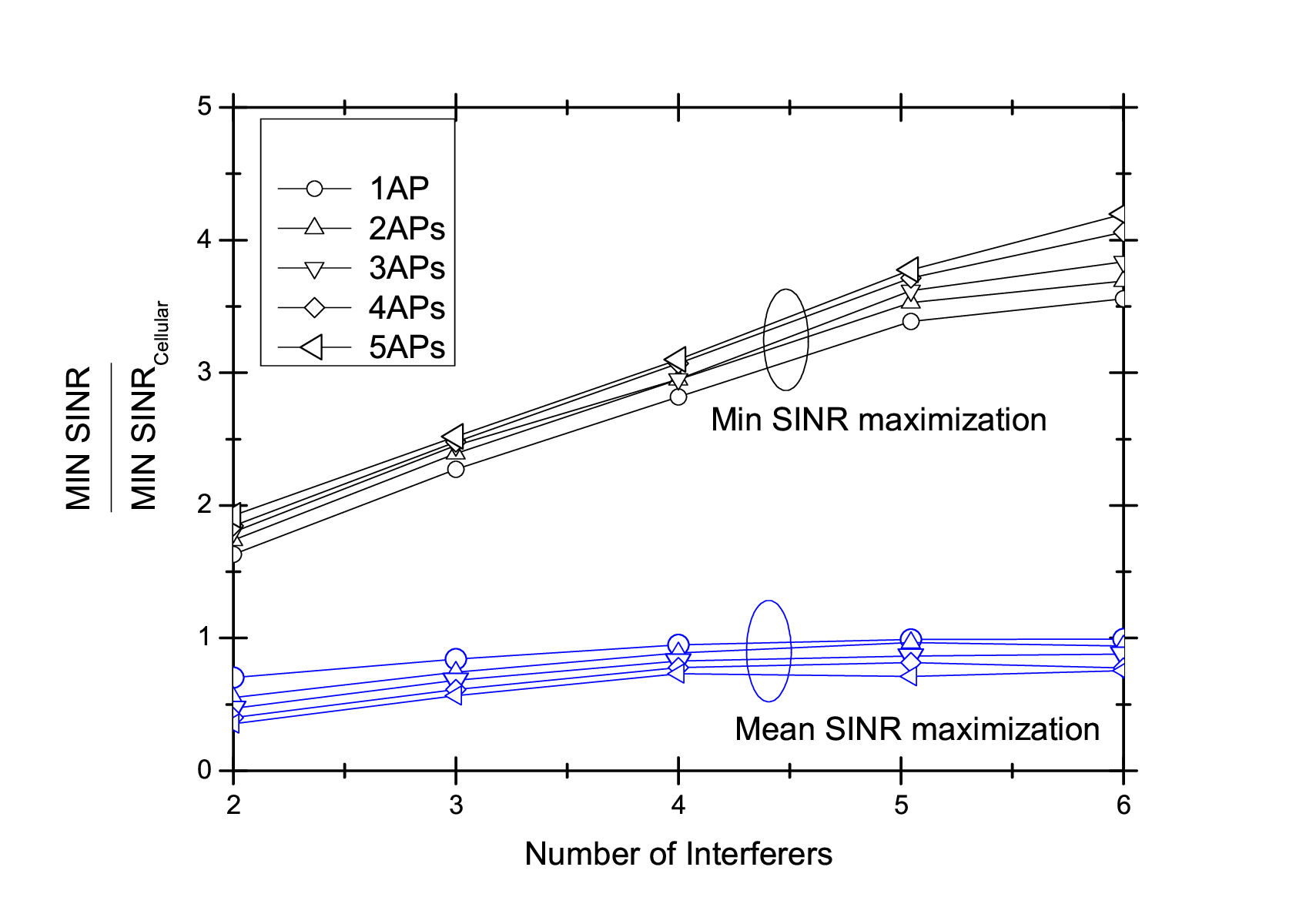}
\caption{The minimum UE SINR gain over the cellular scheme, considering the two optimization problems, i.e., the maximization of the cell's mean SINR and the maximization of the minimum UE SINR.} \label{Fig:Fig_MU_maxmin}
\end{figure}

Regarding the scheme presented in Section V, in Fig.~\ref{Fig:Fig_MU}, the SINR gain of the proposed architecture over the cellular scheme is illustrated, considering both the the optimal solution (i.e., through the exhaustive search) and the sub-optimal one provided by the greedy algorithm (with Q=2). As observed the overall cell's SINR may quadruple compared to the cellular scheme, while the gain diminishes as the number of interferers increases. The greedy solution provides a gain that is close to the corresponding extracted with exhaustive search, for a small number of interferers or APs. Note that the accuracy of the greedy algorithm can be improved by increasing Q, i.e., the optimum topology for Q UEs, at the cost of computational complexity. In Fig.~\ref{Fig:Fig_MU_iter}, the number of searches required by the greedy algorithm (for Q=2) and the exhaustive search approach are depicted. Since, the number of searches is prohibitive for finding the overall SINR for the case of the exhaustive search, the plot in Fig.~\ref{Fig:Fig_MU} is limited to seven co-channel interferes. For more interferers the SINR gain of the hybrid scheme over the cellular is plotted in Fig.~\ref{Fig:Fig_MU_greedy}, using the greedy solution, which is fast enough for optimizing a network with up to 30 co-channel interferes.

The maximization of the cell's average SINR may result in the optimal performance in terms of the overall sum-rate, but it is known that it ignores the fairness among the UEs. To this end, the maximization of the minimum SINR among the UEs is usually employed as the optimization performance criterion. As depicted in Fig.~\ref{Fig:Fig_MU_maxmin}, maximizing the cell's overall SINR greatly affects the individual UEs SINRs. Having the minimum UEs SINR of the cellular scheme as a benchmark, we observe that the greedy solution results in a performance which is worse than that of the cellular scheme, in terms of the maximum minimum UE SINR.

\section{Conclusions}
In this paper, we investigated the performance improvement induced by adopting a multi-RAT communication architecture where the mobile users can be served by either the eNB or a WLAN AP. In the proposed scheme the WLAN APs are wirelessly connected to the eNB and share this broadband connection with other users over WLAN frequencies. Important performance metrics of the proposed hybrid scheme -as the ABEP, average output SINR, ergodic capacity and the OP and the average signal-to-interference-plus noise ratio (SINR)- were theoretically studied and derived in closed form for the single-user case, considering both identically and non-identically fading channels. Furthermore, we considered the case of a multicell MU network aiming to minimize the total inter-cell interference and provided a fast greedy solution to this optimization problem. Additionally, we proposed a novel scheme for achieving frequency reuse one in a single cell and two optimization problems were considered either for maximizing the overall cell's average SINR, or the minimum UE's SINR. A fast greedy algorithm was presented as well, which determines the optimum network topology and provides results close to the optimum ones obtained by exhaustive search. Numerical results showed that the proposed wireless architectures may offer significant performance gains in the presence of multiple interferers, compared to a conventional cellular network.

\renewcommand{\theequation}{A-\arabic{equation}}
\setcounter{equation}{0}
\appendices
\section{Statistics For RVs of the form $C=\frac{A}{1+B}$.}\label{App:derivation}
In this appendix the PDF and CDF of the following RV will be derived
\begin{equation}\label{app:eq1}
C=\frac{A}{1+B}
\end{equation}
with
\begin{equation}
B=\sum_{i=1}^X B_{i}
\end{equation}
where $B_i, A$ are exponential distributed RVs, with parameters $Z_i$ and $Y$, respectively, e.g., the PDF of $A$ is given by
\begin{equation}\label{eq:1}
f_{A}(x)=\frac1{Y}\exp \left( -\frac{x}{Y} \right).
\end{equation}
The RV of the form given in \eqref{app:eq1} represents the output SINR of the $k$th UE in the conventional cellular network, i.e., \eqref{eq:SINR_cell}, as well as for the hybrid Cellular/WLAN, the output SINR of the $k$th UE at the second phase of the indirect communications, i.e., \eqref{eq:SINR_hybrid_indirect_AE}.

\subsection{Non Identical Statistics}
Assuming i.n.d. parameters, the PDF of $B$, can be expressed as follows \cite{Ross}
\begin{equation}\label{eq:pdf_ind_definition}
f_{B}(x)= \left( \prod_{i=1}^{X}\frac{1}{Z_i}\right) \sum_{j=1}^{X} \frac{\exp \left( -\frac{x}{Z_j}\right)}{\prod_{\substack{l=1 \\ l\neq j}}^{X} \left( \frac{1}{Z_l}-\frac{1}{Z_j}\right)}.
\end{equation}
For obtaining the PDF of $C$ the following integral needs to be solved
\begin{equation}\label{eq:7}
f_{C}(\gamma)= \int_0^\infty (1+x) f_{A}\left[(1+x)\gamma\right] f_{B}(x)dx.
\end{equation}
Substituting \eqref{eq:1} and \eqref{eq:pdf_ind_definition} in \eqref{eq:7}, using \cite[eq. (3.381/4)]{Ryzhik} and after some mathematical manipulations yields the closed-form expression of the form A given in Table~II. The corresponding CDF expression for $C$ is given in closed-form in Table~II form A.

\subsubsection{Identical}
Assuming i.i.d. parameters, the PDF of $B$, is a well known chi-square distribution of the form
\begin{equation}\label{eq:2}
\begin{split}
f_{B}(x)&=\frac{Z^{- X}}{\Gamma(X)} x^{X-1} \exp\left( -\frac{x}{Z}\right)
\end{split}
\end{equation}
where $Z=Z_i, \forall i \in [1,X]$ and $\Gamma(\cdot)$ is the Gamma function \cite[eq. (8.310/1)]{Ryzhik}. Substituting \eqref{eq:1} and \eqref{eq:2} in \eqref{eq:7}, and following a similar procedure as the one for the i.n.d. case, $f_{C}(\gamma)$ can be expressed as in Table III form A. The CDF expression for $C$ is given in closed-form in Table III form A.

\renewcommand{\theequation}{B-\arabic{equation}}
\setcounter{equation}{0}
\section{Statistics For RVs of the form $C=\frac{A}{1+B_1+B_2}$.}\label{App:derivation}
In this appendix the PDF and CDF of the following RV will be derived
\begin{equation}\label{app2:eq1}
C=\frac{A}{1+B_{1}+B_{2}}
\end{equation}
with
\begin{equation}
\begin{split}
B_{1}=\sum_{i=1}^{X_1} B_{1,i}, \hspace{0.3cm}
B_{2}=\sum_{i=1}^{X_2} B_{2,i}
\end{split}
\end{equation}
where $B_{{i_1}},B_{{i_2}}, A$ are exponential distributed RVs, with parameters $Z_{i,1},Z_{i,2}$ and $Y$, respectively. The RV of the form given in \eqref{app2:eq1}, for the hybrid Cellular/WLAN network represents the output SINR of the $k$th UE at the direct communication case, i.e., \eqref{eq:SINR_hybrid_direct}, as well as the output SINR of the $k$th UE at the first phase of the indirect communications, i.e., \eqref{eq:SINR_hybrid_indirect_EA}.
Let us define the RV $B=B_{{1}}+B_{{2}}$. The PDF of $B$ can be evaluated using the following integral
\begin{equation}\label{eq:4}
f_{B}(z)=\int_0^z f_{B_{{1}}}(z-x)f_{B_{{2}}}(x)dx.
\end{equation}

\subsection{Non Identical Statistics}
Substituting \eqref{eq:pdf_ind_definition} in \eqref{eq:4} the PDF of $B$, yields the following closed-form expression
\begin{equation}\label{eq:PDF_of_sum_final}
\begin{split}
 f_{B}(z)&= \sum_{j_1=1}^{X_1}\sum_{j_2=1}^{X_2} \frac{\left( \prod_{i=1}^{X_1}\frac{1}{Z_{1,i}}\right)}{\prod_{\substack{k_1=1 \\ k_1\neq j_1}}^{X_1} \left( \frac{1}{Z_{1,k_1}}-\frac{1}{Z_{1,j_1}}\right)}
\\& \times   \frac{\left( \prod_{i=1}^{X_2}\frac{1}{Z_{2,i}}\right)}{\prod_{\substack{k_2=1 \\ k_2\neq j_2}}^{X_2} \left( \frac{1}{Z_{2,{k_2}}}-\frac{1}{Z_{2,j_2}}\right)}\frac{Z_{1,j_1} Z_{2,j_2}}{Z_{1,j_1}-Z_{2,j_2}}\\ & \times
\left[\exp \left( -\frac{z}{Z_{1,j_1}}\right)- \exp \left( -\frac{z}{Z_{1,j_2}}\right)\right].
\end{split}
\end{equation}
In order to derive the PDF of $C$, an integral of the form appearing in \eqref{eq:7} needs to be solved. Substituting \eqref{eq:1} and \eqref{eq:PDF_of_sum_final} in this integral and using \cite[eq. (3.381/4)]{Ryzhik}, the PDF of $C$ is shown in Table II form B.

The corresponding expression for the CDF of $C$ can be obtained by substituting the CDF of $A$, i.e., $F_{A}(x)=1-\exp\left( -\frac{x}{Y}\right)$ and \eqref{eq:PDF_of_sum_final} in the following integral
\begin{equation}\label{eq:CDF_integral_definition}
F_{C}(\gamma)= \int_0^\infty F_{A}[(1+x)\gamma] f_{B}(x)dx.
\end{equation}
The integral in \eqref{eq:CDF_integral_definition} can be solved by using a similar procedure as that used for deriving the PDF of $C$ (shown in Table II form B) and thus after some mathematical manipulations a closed-form expression for the CDF of $C$ can be obtained, which is given in Table II form B.

\subsection{Identical Statistics}
For the i.i.d. case, substituting \eqref{eq:2} in \eqref{eq:4} the PDF of $B$ becomes
\begin{equation}\label{eq:5}
\begin{split}
f_{B}&(z)=\frac{1}{Z_2^{X_2} \Gamma\left(X_2\right)}\frac{1}{Z_1^{X_1} \Gamma\left(X_1\right)}\\ & \times  \int_0^z \left( z-x\right)^{X_2-1}   x^{X_1-1} \\ & \hspace{1cm}\times\exp \left( -\frac{z-x}{Z_2}\right) \exp \left( -\frac{x}{Z_1}\right)dx.
\end{split}
\end{equation}
Similar type of integrals have been also identified in \cite{6512535}. By employing the binomial identity and using \cite[eq. (3.351/1)]{Ryzhik}, an alternative to the solution provided in \cite{6512535} can be provided as
\begin{equation}\label{eq:81}
\begin{split}
f_{B}&(z)=\frac{1}{Z_2^{X_2} \Gamma\left(X_2\right)}\frac{\exp \left( -\frac{z}{Z_2}\right)}{Z_1^{X_1} \Gamma\left(X_1\right)}   \\ &
\times \sum_{j=0}^{X_2-1} \binom{X_2-1}{j} \frac{z^{X_2-1-j}(-1)^j}{\left( \frac1{Z_1}-\frac1{Z_2}\right)^{j+X_1}} \\ & \times \left[ \Gamma\left(j+N\right)-\Gamma\left[j+X_1, \left( \frac1{Z_1}-\frac1{Z_2}\right)z\right]\right].
\end{split}
\end{equation}
Substituting \eqref{eq:1} and \eqref{eq:81} in \eqref{eq:7} and using the finite series representation for the incomplete Gamma function \cite[eq. (06.06.06.0005.01)]{wolfram}, \cite[eq. (3.381/4)]{Ryzhik} and after some straight forward mathematical manipulations the PDF of $C$ is shown at Table III form B.

Furthermore, based on \eqref{eq:CDF_integral_definition} and using a similar procedure as that used for deriving the PDF of $C$, a closed-form expression for the CDF of $C$ is shown at Table III form B.

\balance

\end{document}